\documentclass[english,showpacs,amsmath,twocolumn,prb]{revtex4}
\usepackage[T1]{fontenc}
\usepackage[latin9]{inputenc}
\usepackage{textcomp}
\usepackage{amstext}
\usepackage{graphicx}

\makeatletter

\providecommand{\tabularnewline}{\\}

\@ifundefined{textcolor}{}
{%
 \definecolor{BLACK}{gray}{0}
 \definecolor{WHITE}{gray}{1}
 \definecolor{RED}{rgb}{1,0,0}
 \definecolor{GREEN}{rgb}{0,1,0}
 \definecolor{BLUE}{rgb}{0,0,1}
 \definecolor{CYAN}{cmyk}{1,0,0,0}
 \definecolor{MAGENTA}{cmyk}{0,1,0,0}
 \definecolor{YELLOW}{cmyk}{0,0,1,0}
 }

\@ifundefined{showcaptionsetup}{}{%
 \PassOptionsToPackage{caption=false}{subfig}}
\usepackage{subfig}
\makeatother

\usepackage{babel}

\begin{document}

\title{Theory of Linear Optical Absorption in B$_{\text{12}}$ Clusters:
Role of the geometry}

\author{Sridhar Sahu and Alok Shukla}

\affiliation{Department of Physics, Indian Institute of Technology, Bombay, Powai,
Mumbai 400076, INDIA}

\email{sridhar@phy.iitb.ac.in, shukla@phy.iitb.ac.in}
\begin{abstract}
Boron clusters have been widely studied theoretically for their geometrical
properties and electronic structure using a variety of methodologies.
An important cluster of boron is the B$_{12}$ cluster whose two main
isomers have distinct geometries, namely, icosahedral ($I_{h}$) and
quasi planar ($C_{3v}$). In this paper we investigate the linear
optical absorption spectrum of these two B$_{12}$ structures with
the aim of examining the role of geometry on the optical properties
of clusters. The optical absorption calculations are performed using
both the semi-empirical and the \emph{ab initio} approaches. The semi-empirical
approach uses a wave function methodology employing the INDO model
Hamiltonian, coupled with large-scale configuration interaction (CI)
calculations, to account for the electron-correlation effects. The
\emph{ab initio} calculations are performed within a time-dependent-density-functional-theory
(TDDFT) methodology. The results for the two approaches are in very
good qualitative agreement with each other. Quantitatively speaking,
results agree with each other in the lower energy region, while in
the higher energy region, features predicted by the TDDFT approach
are red-shifted as compared to the INDO-CI results. Both the approaches
predict that the optical absorption begins at much lower energies
in the icosahedral cluster as compared to the planar one, a fact which
can be utilized in experiments to distinguish between the two geometries.
At higher energies, both the isomers exhibit plasmon-like excitations. 
\end{abstract}


\maketitle

\section{Introduction}

\label{sec:intro}

In the recent years, the study of the boron clusters has attracted
considerable attention due to their potential applications as hydrogen
storage devices,\cite{ozturk,McKee} hard semiconducting solids and
various other properties.\cite{Mutterties,lipscomb} Extensive theoretical
studies of one-, two-, and three-dimensional boron clusters of different
sizes have been carried out by various researchers.\cite{kbamba,kawai,kato,boustani-1,boustani-2,fujimori,hayami,zhai,b12-ionicity-prl,atis,jemmis-prl}
Of these, icosahedral B$_{\text{12}}$ cluster\cite{Mutterties} has
generated a lot of interest in recent years possibly because of its
high symmetry and its occurrence as the fundamental structural unit
in crystalline boron, and various boron-rich solids.\cite{hubert,perkin}In
an early work Bambakidis and Wagner\cite{kbamba} computed the structural
and cohesive properties of icosahedral B$_{12}$ using the SCF-X$\alpha$-SW
approach, and estimated its HOMO-LUMO gap. Kawai and Ware\cite{kawai}
employed Car-Parinello \emph{ab initio} molecular dynamics approach
to analyze the structural instabilities of cage-like B$_{\text{12}}$.
Kato and Yamashita\cite{kato} performed \emph{ab initio} Hartree-Fock
calculations to optimize the geometries of various neutral and cationic
boron clusters, and obtained a triplet ground state with a trigonal
bipyramid structure for the B$_{\text{12}}$ cluster. Boustani,\cite{boustani-1,boustani-2}
using both \emph{ab initio} density-functional theory (DFT),\cite{boustani-1}
and wave function based quantum-chemical\cite{boustani-2} approaches,
presented detailed studies of various boron clusters, including several
structural isomers of B$_{12}$. Using a semiempirical approach, Fujimori
and Kimura\cite{fujimori} explored the nature of bonding in icosahedral
clusters of group III atoms, including cage-like B$_{\text{12}}$.
Hayami,\cite{hayami} using an \emph{ab initio} DFT based approach
studied the encapsulating properties of the icosahedral B$_{\text{12}}$.
 Zhai \emph{et al.,\cite{zhai}} in a recent joint theoretical and
experimental study investigated some small boron clusters, including
B$_{12}$, and argued that these clusters prefer planar aromatic structures.
In a first-principles study, He \emph{et al.},\cite{b12-ionicity-prl}
explored the ionicities of boron-boron bond in icosahedral B$_{12}$
caused by symmetry breaking. Atis \emph{et al.},\cite{atis} based
upon an \emph{ab initio} DFT methodology, presented a theoretical
study of several neutral boron clusters, along with an analysis of
relative stability of various isomers, including B$_{12}$ isomers.
 Prasad and Jemmis\cite{jemmis-prl} in a first principles study examined
the structure of several large boron clusters and concluded that the
ones constructed from the B$_{\text{12}}$ icosahedron as the basic
unit are more stable as compared to the fullerene-like structures. 

In spite of several studies of structural properties of boron clusters,
very few studies of their dielectric response properties exist. For
example, Reis \emph{et al}.,\cite{reis-1,reis-2} computed the static
linear and nonlinear optical susceptibilities of rhombic B$_{\text{4}}$
clusters using a first-principles methodology. In an earlier work,
by means of \emph{ab initio} correlated calculations, we had computed
the static dipole polarizabilities of ladder-like boron clusters.\cite{ayj}
Although optical absorption spectra of icosahedral clusters of atoms
such as Al and Pb have recently been computed,\cite{xie} to the best
of our knowledge, no such calculations for the linear or non-linear
optical response of B$_{\text{12}}$ clusters (quasi-planar or icosahedral)
have been performed. Given the fact that B$_{\text{12}}$ clusters
were recently discovered experimentally,\cite{zhai} it is of considerable
interest to compute their optical properties. Theoretical calculations
of optical absorption spectra, coupled with the experimental measurements,
can be used to identify clusters of various shapes and sizes, and
to distinguish between various isomers. With this aim in mind, we
present a theoretical study of the frequency-dependent linear optical
response of B$_{\text{12}}$ icosahedral cluster, as well as its quasi-planar
isomer which was recently discovered experimentally by Zhai \emph{et
al.}\cite{zhai} Such comparative studies are important because they
also help us in understanding the role of geometry in the optical
response of clusters. Our calculations have been performed using both
a semiempirical INDO\cite{INDO} model based approach within the configuration-interaction
(CI) framework, as well as an \emph{ab initio} time-dependent density
functional theory (TDDFT) based methodology.\cite{tddft-1,tddft-2}
Linear optical absorption spectra computed using the two approaches
are in full qualitative agreement with each other. Regarding quantitative
aspects, agreement in the lower energy region is very good, while
the TDDFT results are redshifted as compared to the INDO-CI results
in the high energy region. The predictions of our calculations can
be tested in future optical absorption experiments on these clusters.

Remainder of the paper is organized as follows. In the next section
we present the theoretical aspects of our approach, to be followed
by the presentation and discussion of our results in section \ref{sec:results}.
Finally, in section \ref{sec:conclusion} we present the conclusions

\section{Theoretical Background}

\label{sec:theory}

For the present study, we adopted a wave-function based electron-correlated
methodology employing the semi-empirical intermediate-neglect of differential
overlap (INDO) model developed by Pople and coworkers.\cite{INDO,pople-book}
The INDO model, like its predecessor complete neglect of differential
overlap (CNDO) model,\cite{CNDO} employs an effective valence-electron
Hamiltonian which uses Slater-type-orbital (STO) as basis functions,
and several of its one- and two-electron integrals are approximated
using a semi-empirical parameterization scheme. The main difference
between the INDO and the CNDO models is that, in the INDO model one-center
Coulomb repulsion integrals include more nonzero terms as compared
to the CNDO model, leading generally to a better description of the
excited state energies. For a detailed discussion of the theory behind
these approaches, and essential differences therein, we refer the
reader to the book by Pople and Beveridge.\cite{pople-book}

We adopted a semi-empirical approach, as against a fully \emph{ab
intio} one, because, with twelve atoms, an \emph{ab initio} correlated
calculation with even modest-sized basis sets becomes intractable.
On the other hand, the INDO method, with its smaller basis set (four
basis functions per atom), allows one to treat electron-correlation
effects at a much higher level than what is feasible using an \emph{ab
initio} approach. Our calculations are initiated at the Hartree-Fock
(HF) level, within the INDO model, using a computer program developed
recently by us.\cite{shukla} The INDO-HF molecular orbitals (MOs),
are used to transform the Hamiltonian from the original atomic-orbital
(AO) to the MO representation, which is subsequently used in the post-HF
correlated calculations. 

There are several variants of the INDO approach in vogue which differ
from each other in terms of the semi-empirical parameters used. In
this work we have used the original INDO parameterization proposed
by Pople and coworkers.\cite{INDO} For the calculations of spectroscopic
properties such as the excitation energies, CNDO/S\cite{CNDO-S} and
the INDO/S\cite{INDO-S1,INDO-S2,INDO-S3} approaches have been used
so frequently that it is virtually impossible to cite all of them.
INDO/S method was parameterized by Zerner and coworkers\cite{INDO-S1,INDO-S2,INDO-S3}
to reproduce spectra of small aromatic molecules with the use of low-order
CI approaches, such as the singles-CI (SCI) method.  However, when
Adachi and Nakamura benchmarked the performance of the INDO/S method
for the case of organic dyes, they found that the excitation energies
obtained were significantly blue-shifted as compared to the experimental
results.\cite{INDOS-dye} This led the authors to conclude that while
the INDO/S method performs well for the ultraviolet region of the
spectrum possibly because it was parameterized for small aromatic
molecules, its performance is not all that good in the visible region
of the spectrum.\cite{INDOS-dye} Therefore, we decided to use the
original INDO parameters,\cite{INDO} coupled with a high-level correlation
scheme (see below) so that the influence of parameters is neutralized
to a certain extent. Indeed, recently El-Shahawy \emph{et al.\cite{indo-paracetamol}}
also used this original INDO approach\cite{INDO} to calculate the
excitation energies of the paracetamol molecule successfully. However,
to further benchmark our INDO-CI approach, as well as to approach
the problem from a complementary perspective, we have also performed
calculations of the optical absorption spectra of the two isomers
using the \emph{ab initio} TDDFT approach.\cite{tddft-1,tddft-2}
This, we believe, has helped us tremendously in critically analyzing
our INDO-CI results.

The correlated calculations, beyond the INDO-HF, are performed using
the multi-reference singles-doubles configuration-interaction (MRSDCI)
approach as implemented in the computer program MELD.\cite{meld}
MRSDCI approach is a well-established quantum-chemical approach in
which one considers singly- and doubly-excited configurations from
a number of reference configurations, leading to a good treatment
of electron correlations both for the ground and the excited states,
in the same calculation. Using the ground- and excited-state wave
functions obtained from the MRSDCI calculations, electric dipole matrix
elements are computed and subsequently utilized to compute the linear
absorption spectrum assuming a Lorentzian line shape. By analyzing
the wave functions of the excited states contributing to the peaks
of the computed spectrum obtained from a given calculation, bigger
MRSDCI calculations are performed by including a larger number of
reference states. The choice of the reference states to be included
in a given calculation is based upon the magnitude of their contribution
to the CI wave function of the excited state (or states) contributing
to a peak in the spectrum. This procedure is repeated until the computed
spectrum converges within an acceptable tolerance, and all the configurations
contributing significantly to various excited states are included
in the list of the reference states. In the past, we have used such
an iterative MRSDCI approach on a number of conjugated polymers to
perform large-scale correlated calculations of their linear and nonlinear
optical properties.\cite{mrsd-calc}

\section{Calculations and Results}

\label{sec:results}

\subsection{Geometries }

\label{sub:geometry}

\begin{figure}
\subfloat[]{\includegraphics[width=3.5cm]{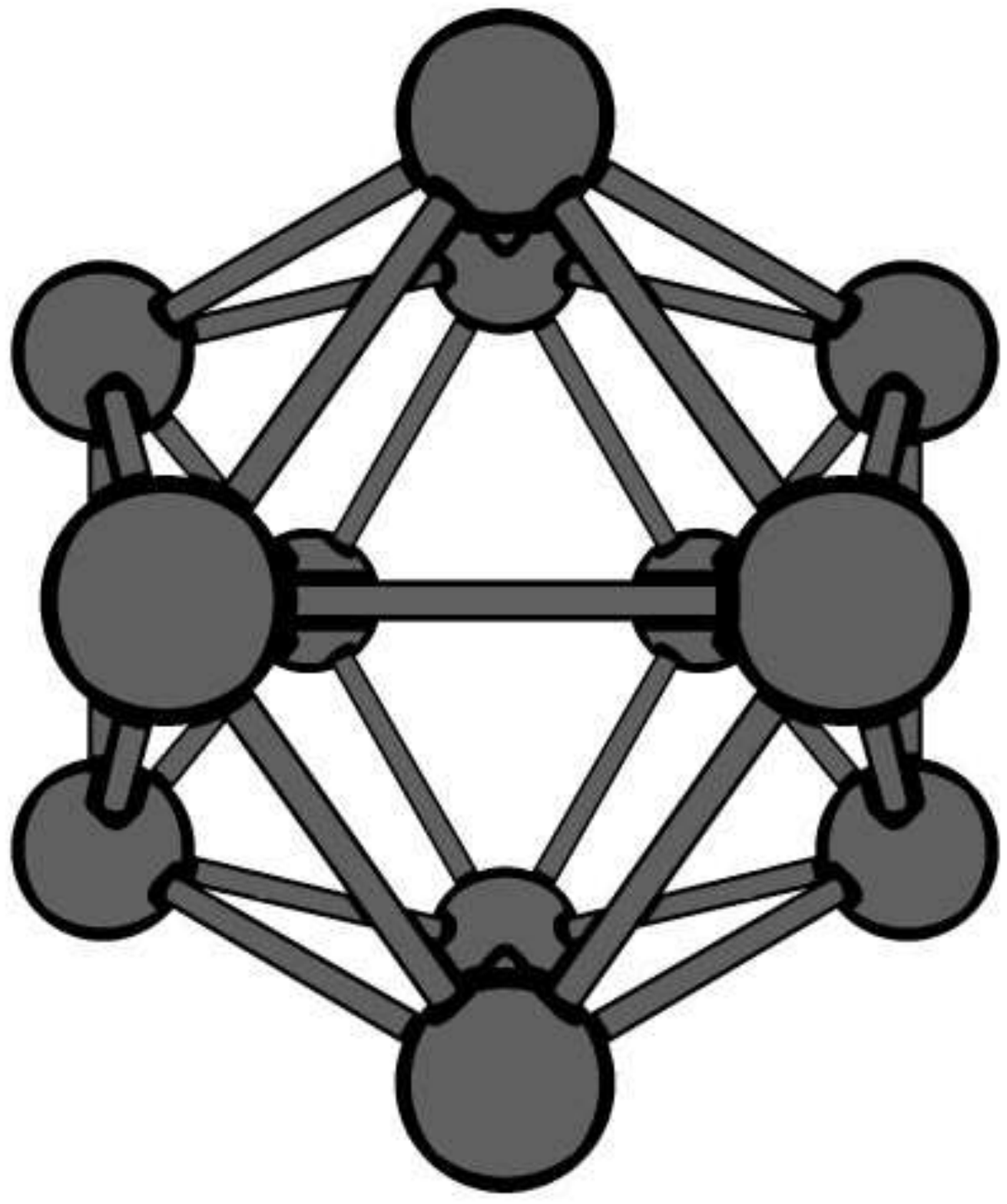}

\label{fig:icos}}\subfloat[]{\includegraphics[width=3.5cm]{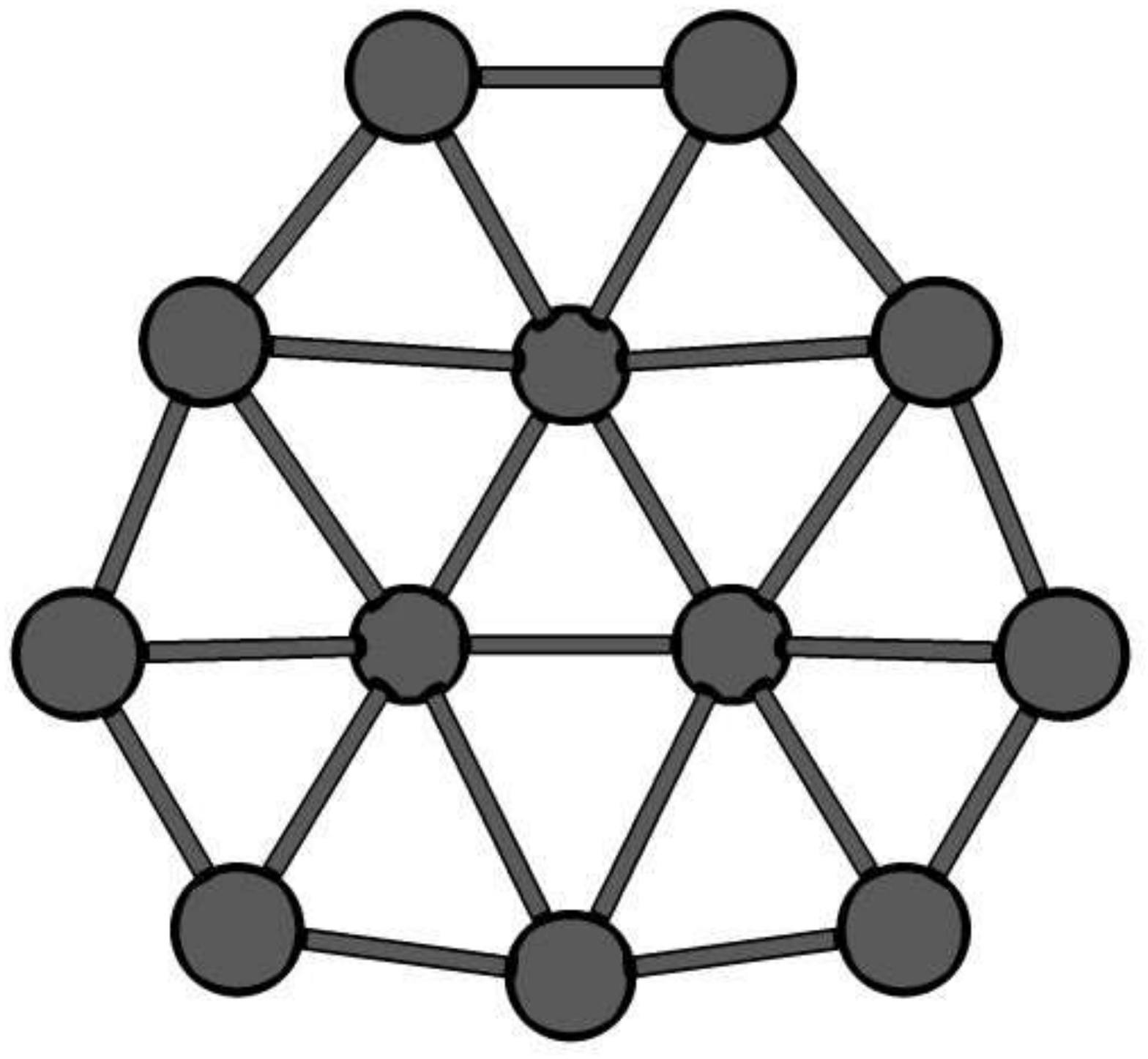}

\label{fig:planar}}

\caption{ Structures of: (a) icosahedral ($I_{h}$) and, (b) quasi-planar ($C_{3v}$)
clusters of B$_{12}$ considered in these calculations. }

\label{Fig:struct} 
\end{figure}

In Figs. \ref{fig:icos} and \ref{fig:planar} we depict the structures
of the icosahedral ($I_{h}$) and quasi-planar ($C_{3v}$) isomers
of B$_{12}$, respectively. The ground state optimized geometries
of both the isomers of B$_{\text{12}}$ were obtained at the INDO-HF
level. For the icosahedron we obtained the edge length to be 1.702$\;$Å
which is virtually identical to that obtained by Hayami for the same
structure.\cite{hayami} In order to further ensure that our geometry
was reasonable, using the Gaussian03\cite{gaussian} program we also
performed geometry optimization for both the isomers within the DFT
approach, employing the B3LYP functional, and a 6-31g(d) basis set.
This calculation also yielded the edge-length of 1.70 Å for the $I_{h}$
structure, in perfect agreement with our INDO-HF results, and those
of Hayami.\cite{hayami} For the quasi-planar structure of $C_{3v}$
symmetry, we used the geometry optimized by Boustani with the bond
lengths 1.60 Å, 1.64 Å, and 1.65 Å.\cite{boustani-2} Our own B3LYP/6-31g(d)
optimization for this system yields corresponding bond lengths to
be 1.63 Å, 1.66 Å, and 1.68 Å, again in good agreement with the results
of Boustani.\cite{boustani-2} However, our results both at the INDO-HF
and the INDO-CI level predict cage structure to be more stable as
compared to the planar one, which is exactly opposite to the result
obtained by Boustani.\cite{boustani-2} In a recent experimental study
Zhai \emph{et al.}\cite{zhai} discovered the quasi-planar B$_{12}$
isomer, without ruling out the possibility that the larger planar
boron clusters will eventually fragment into B$_{12}$ icosahedra.
The purpose of our work, in any case, is not to examine the relative
stability of the two isomers, but to investigate their linear optical
response.

\subsection{Molecular Orbitals}

\label{sub:MOs}

\begin{figure}
\subfloat[HOMO-2]{\includegraphics[angle=90,scale=0.15]{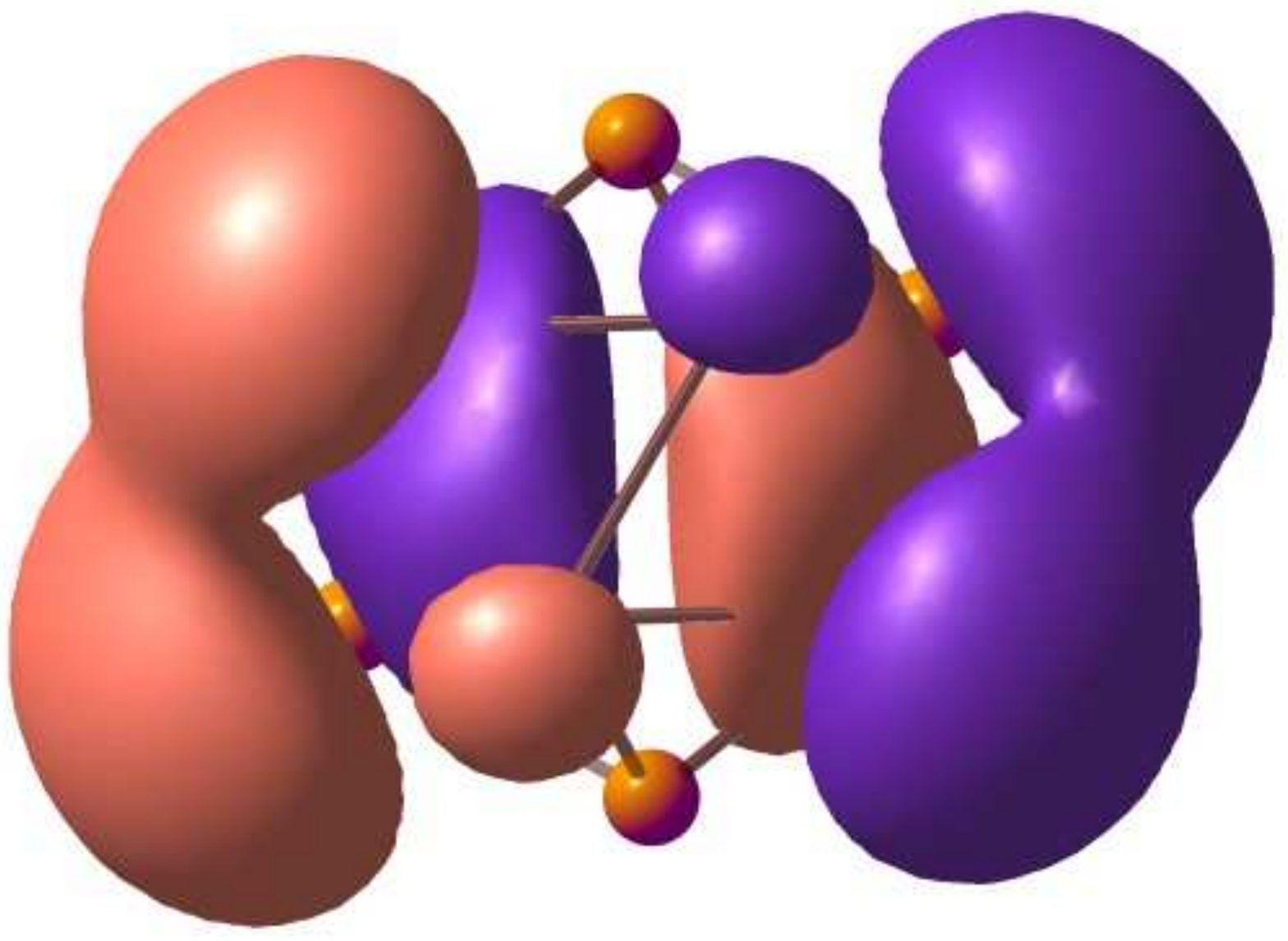}

}\subfloat[HOMO-1]{\includegraphics[scale=0.15]{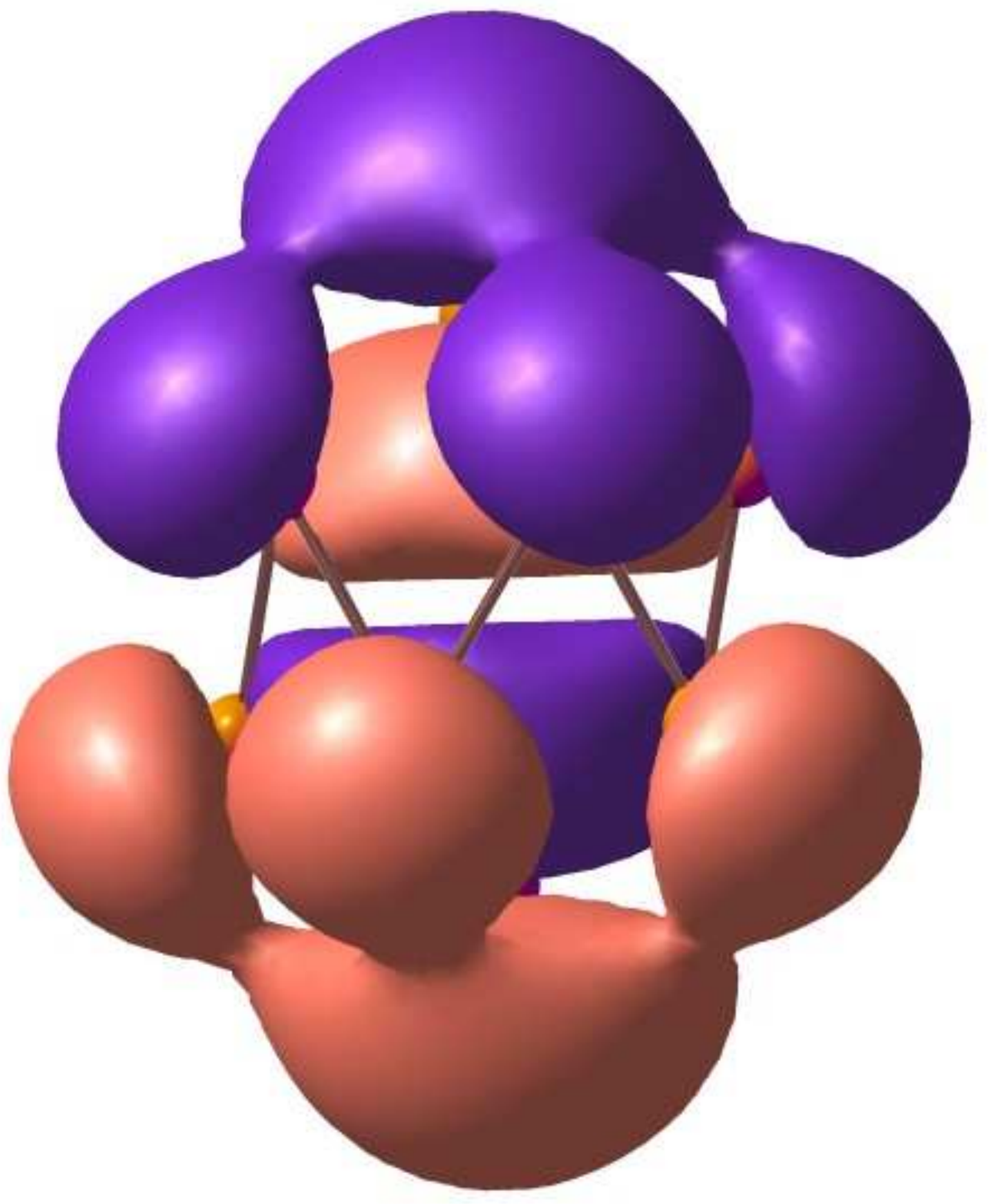}

}\subfloat[HOMO]{\includegraphics[scale=0.15]{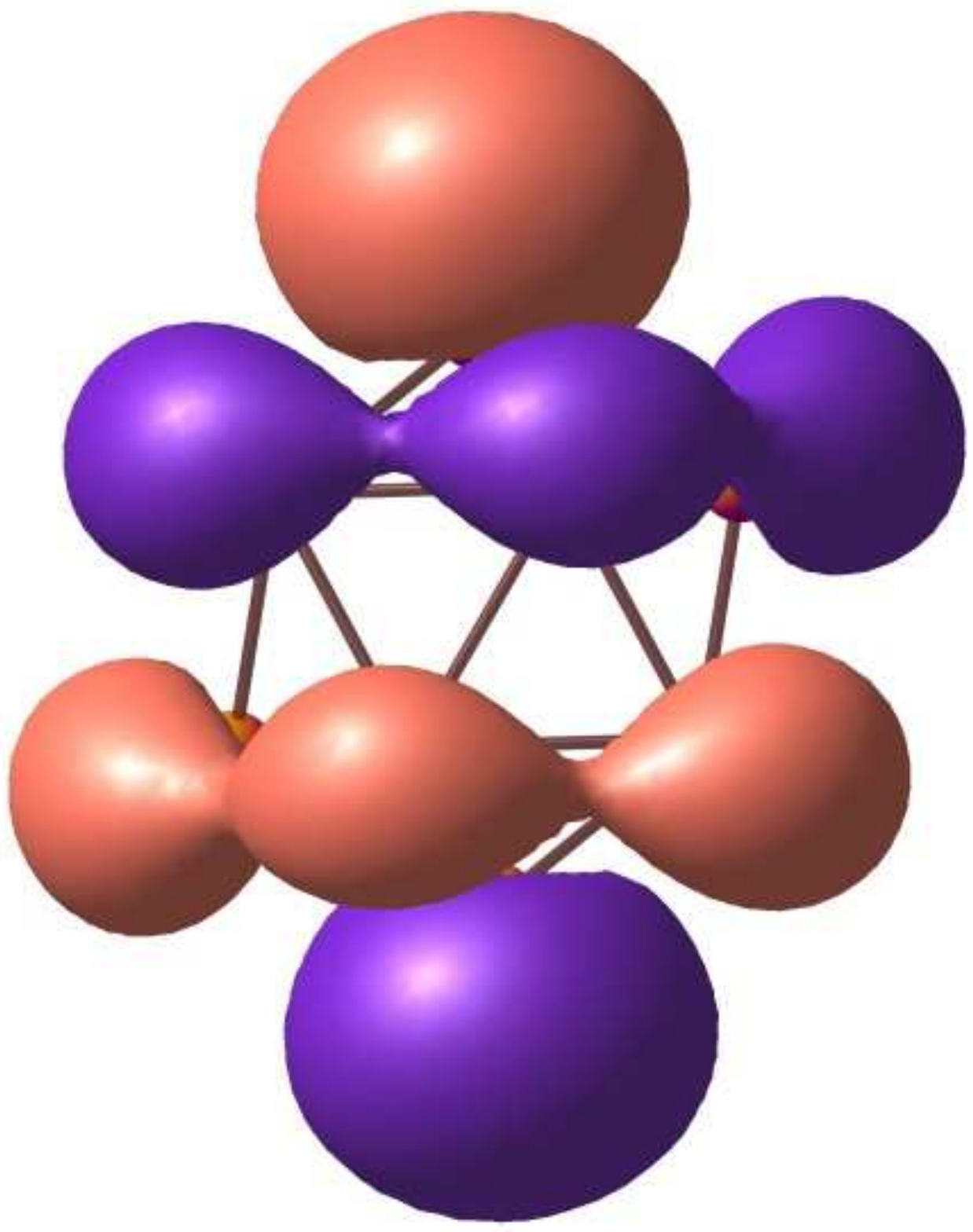}

\label{fig:indo-homo}}

\subfloat[LUMO]{\includegraphics[scale=0.15]{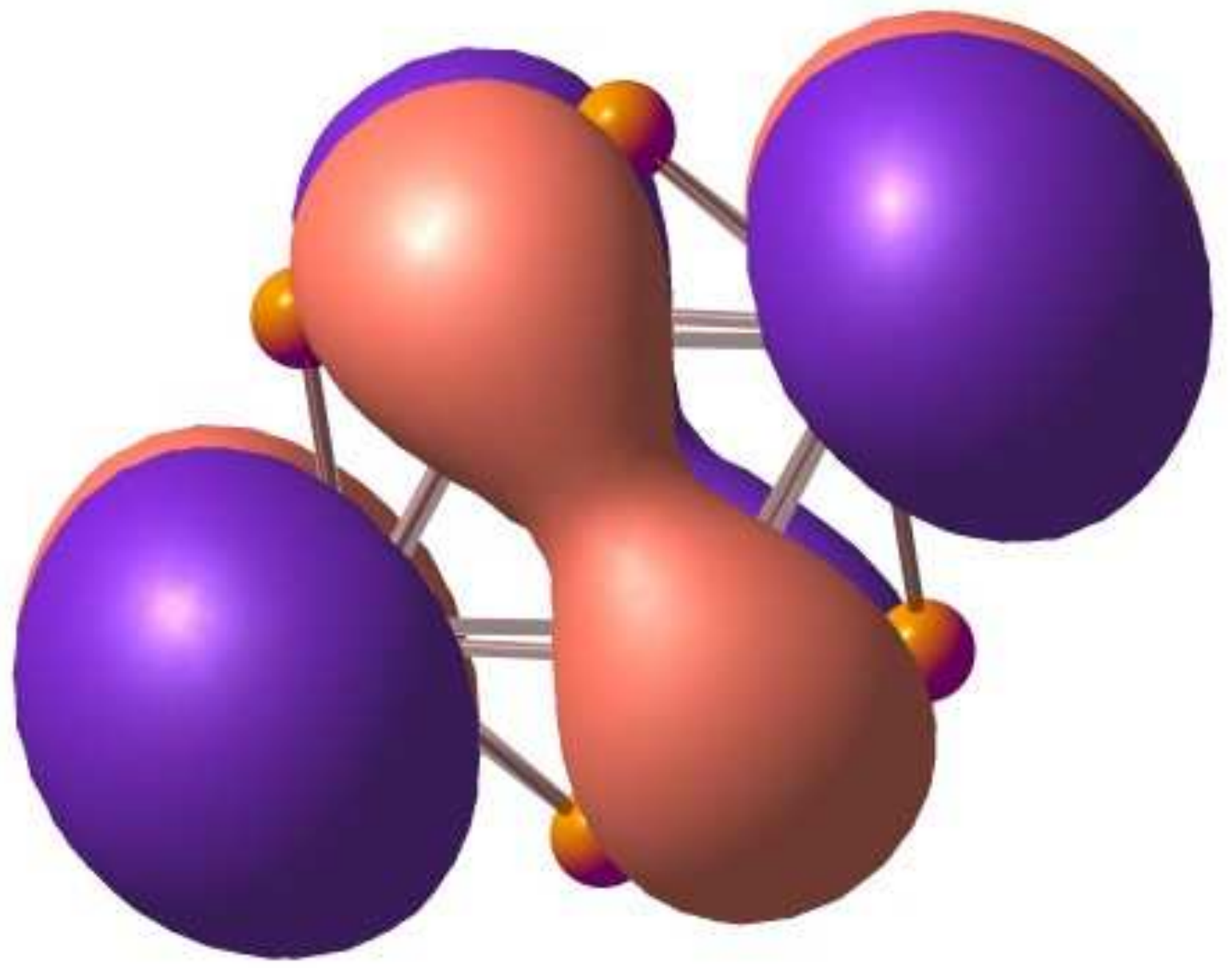}

}\subfloat[LUMO+1]{\includegraphics[scale=0.15]{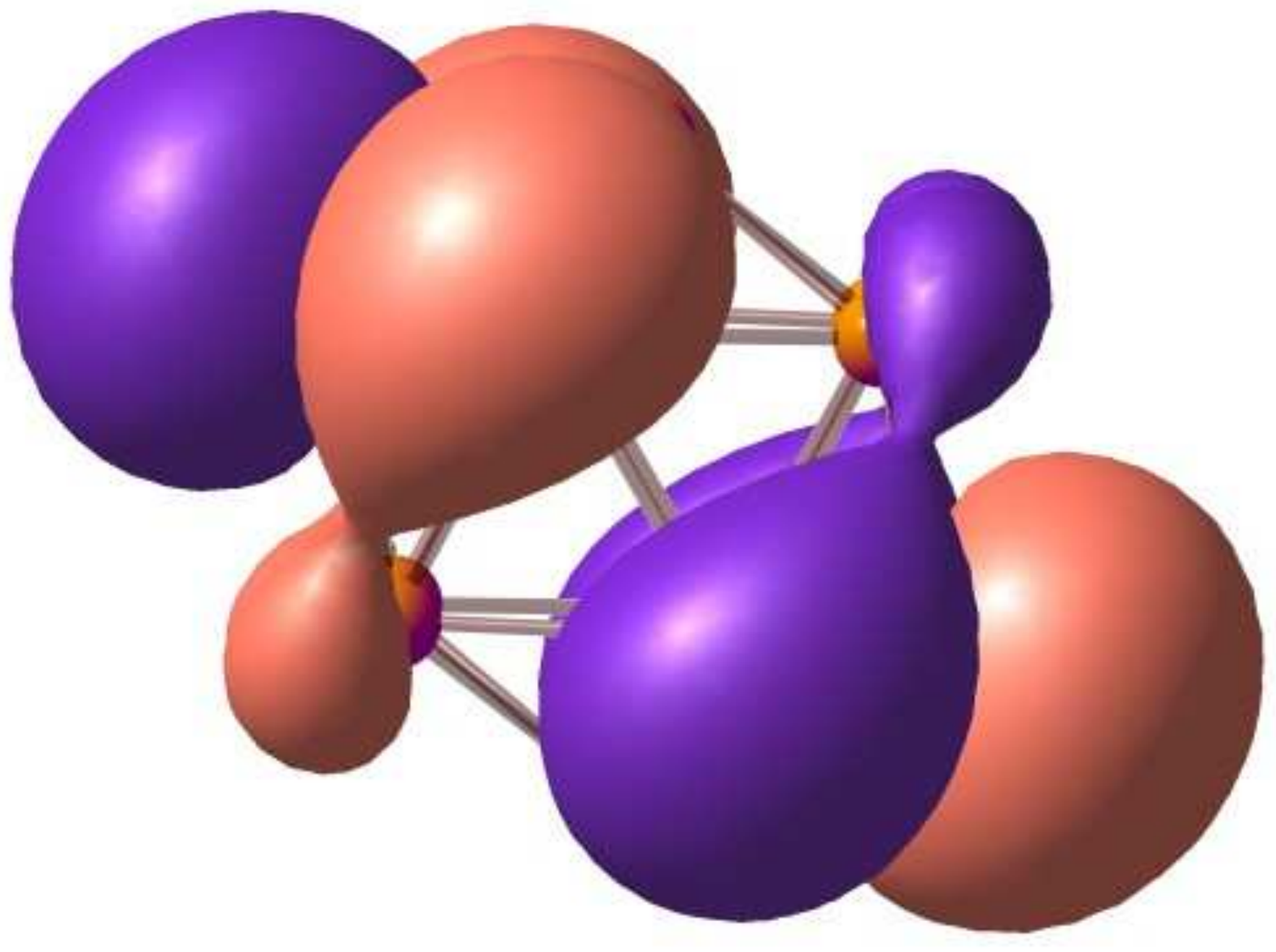}

}\subfloat[LUMO+2]{\includegraphics[scale=0.15]{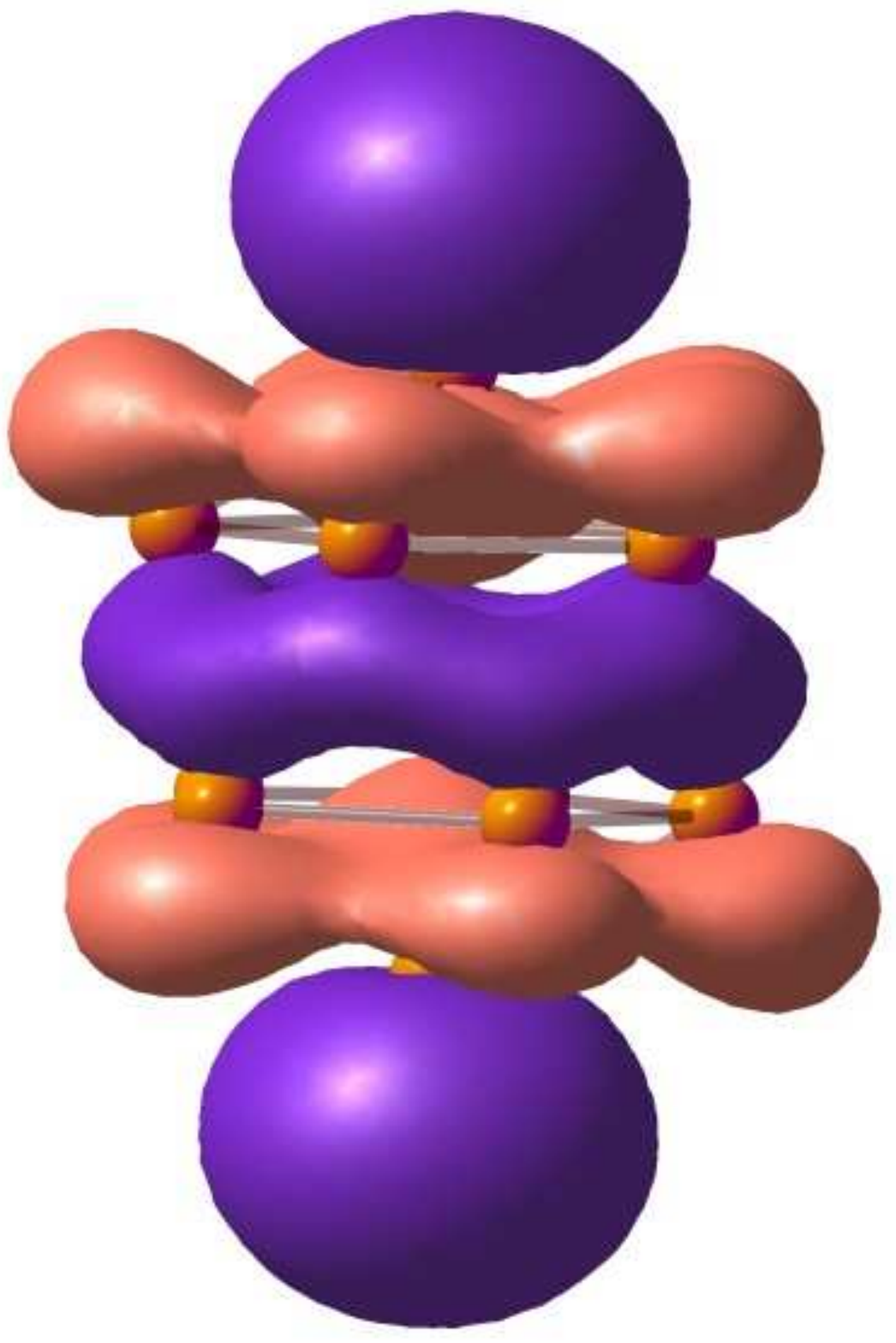}

\label{fig:indo-lumo+2}}

\caption{(Color online) Molecular orbitals (iso plots) of icosahedral B$_{\text{12}}$
obtained from INDO-HF calculations.}

\label{fig:indo-molorb-icos}
\end{figure}

\begin{figure}
\subfloat[HOMO-1]{\includegraphics[scale=0.15]{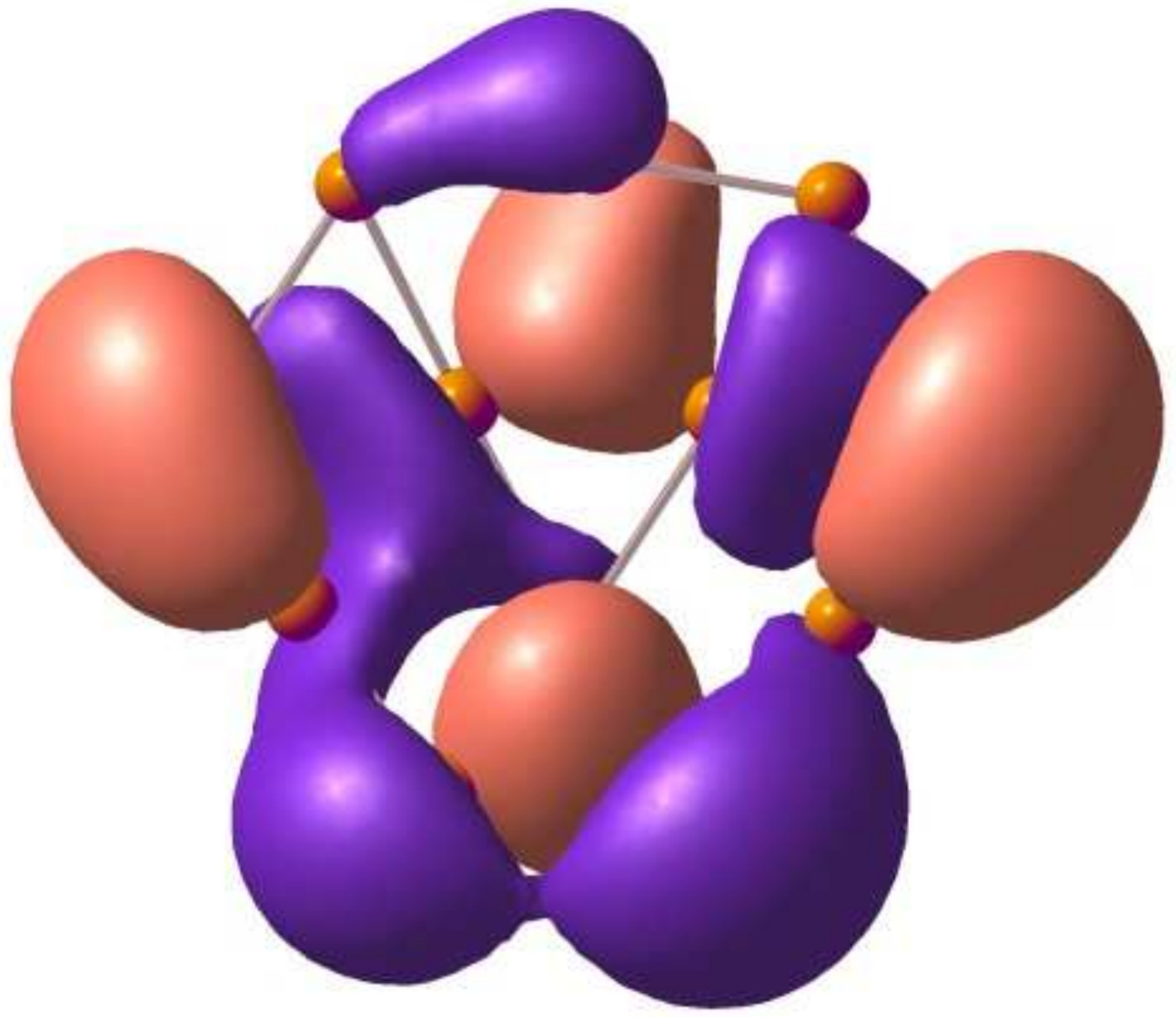}

}\subfloat[HOMO]{\includegraphics[scale=0.15]{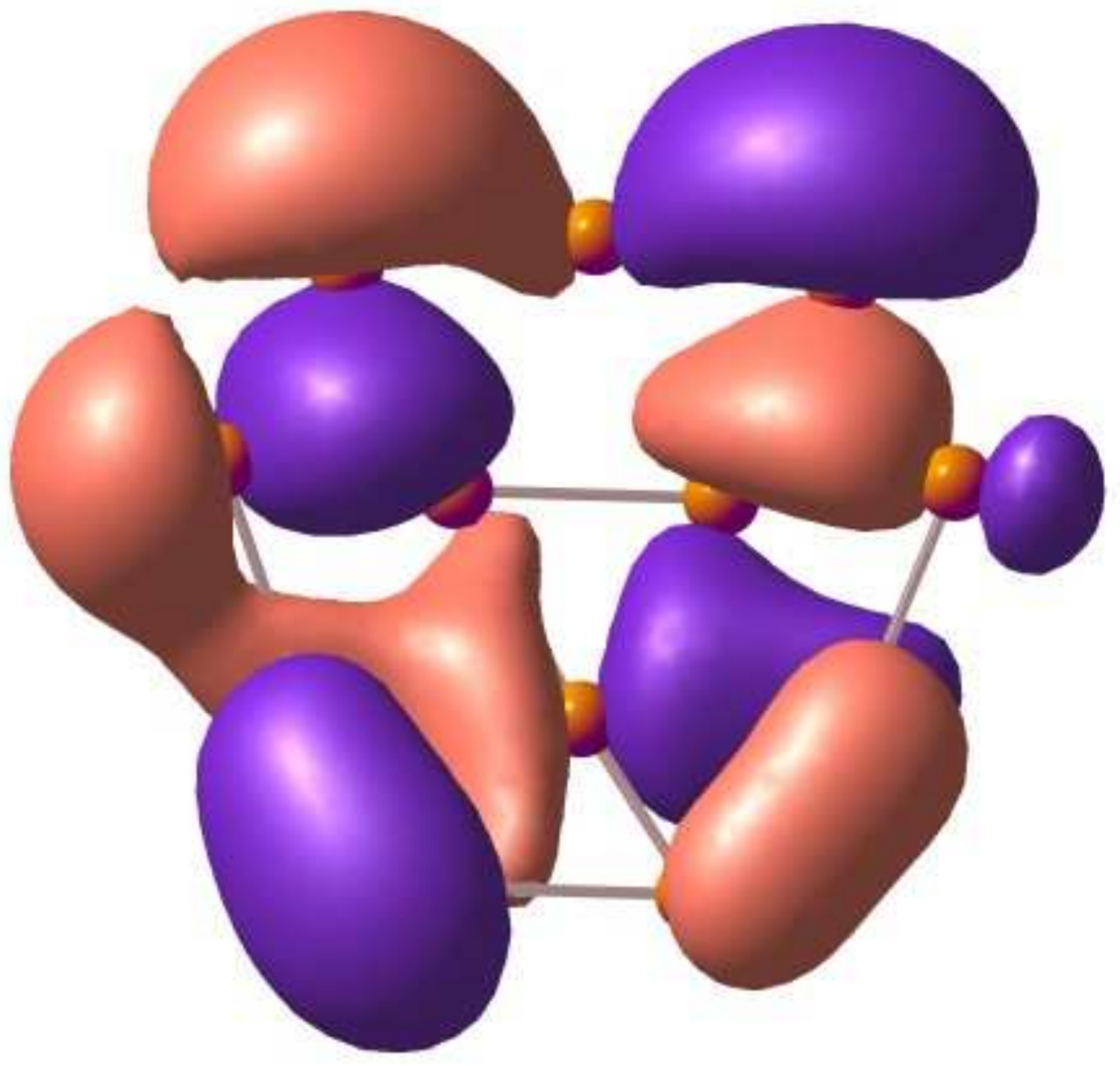}

}\subfloat[LUMO]{\includegraphics[scale=0.15]{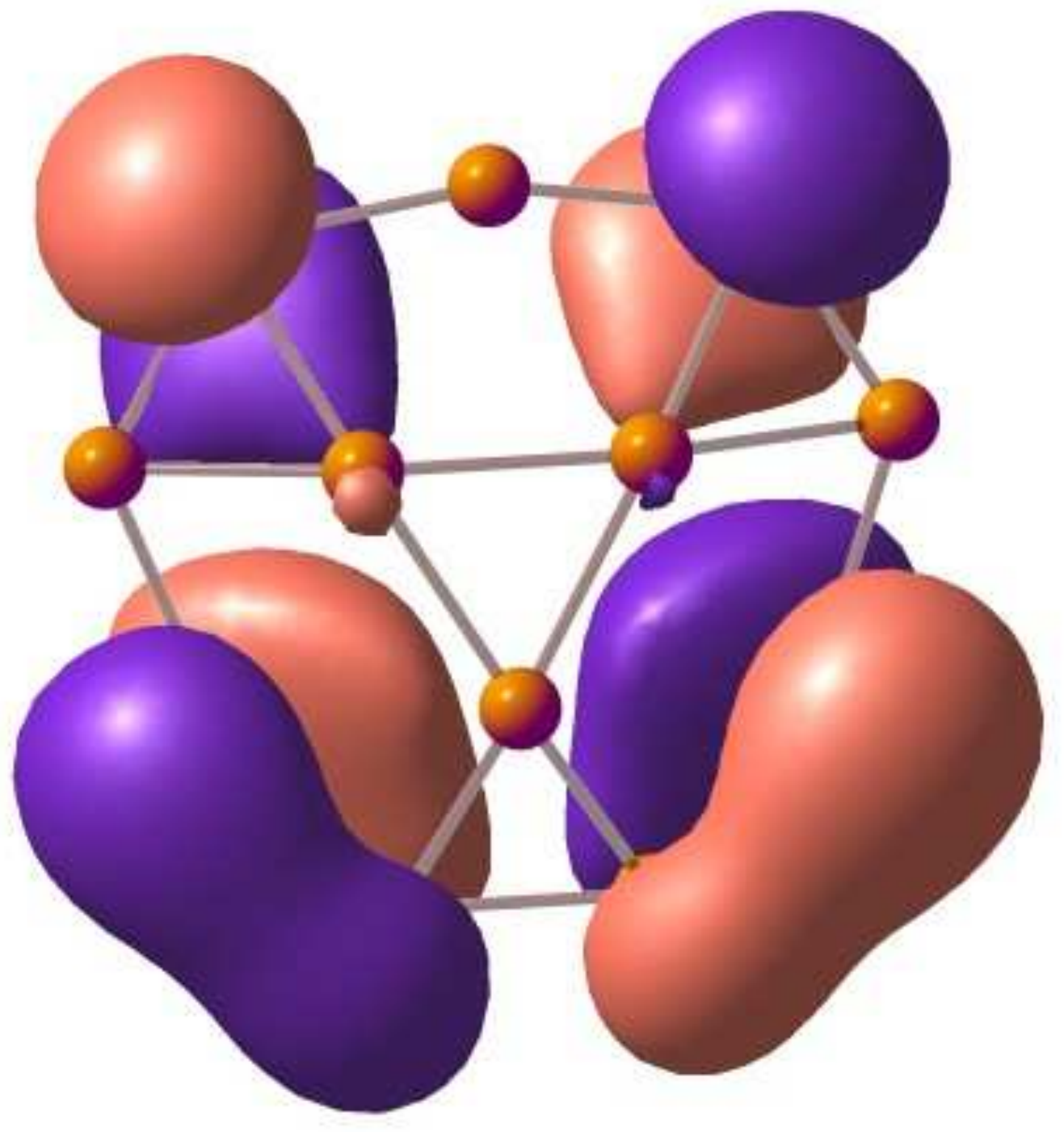}

}

\subfloat[LUMO+1]{\includegraphics[scale=0.15]{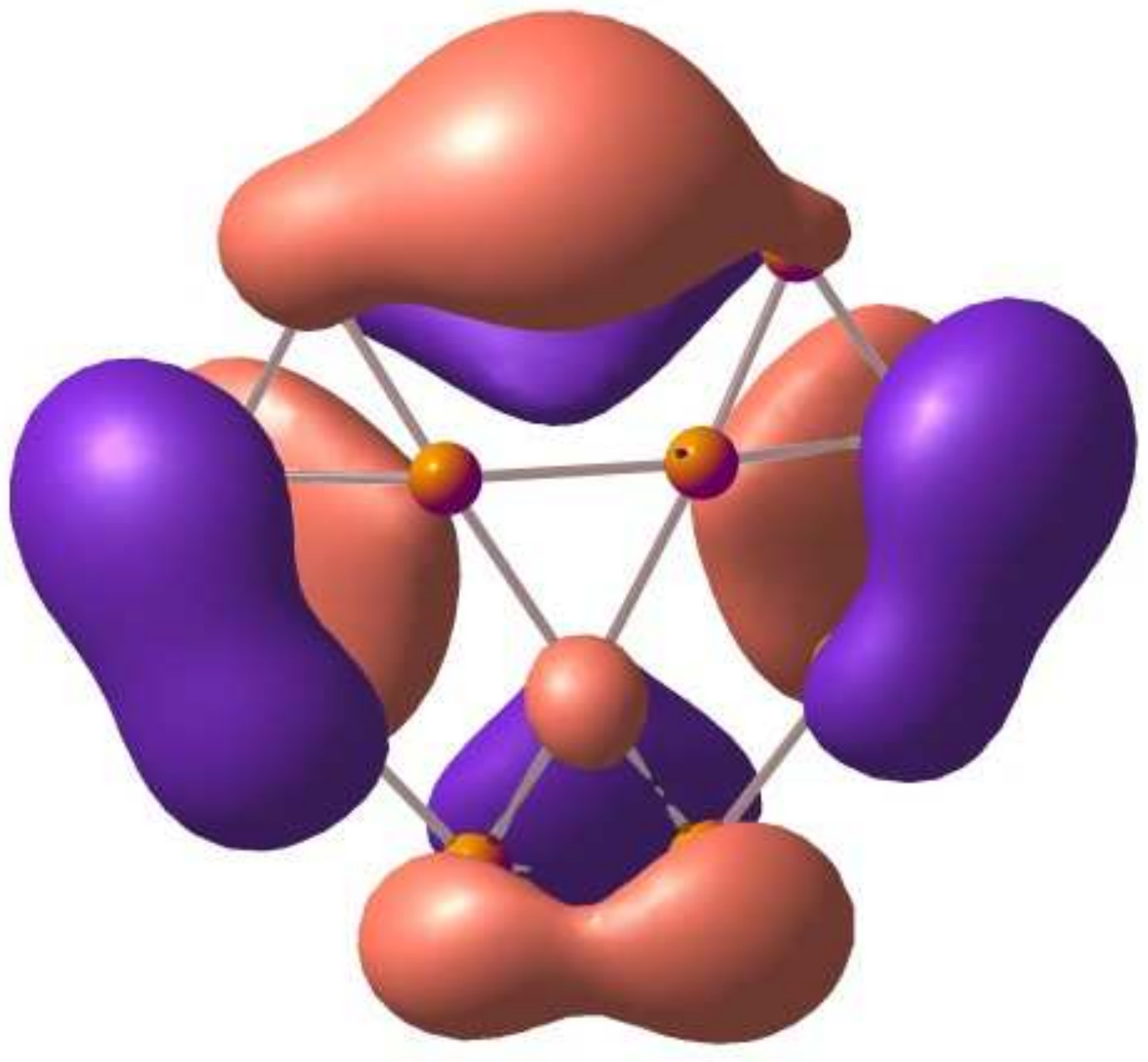}

}

\caption{(Color online) Molecular orbitals (iso plots) of quasi-planar B12
obtained from the INDO-HF calculations.}

\label{Fig:indo-molorbs-planar}
\end{figure}

Next we present the plots of the molecular orbitals (MOs) of both
the icosahedral and quasi-planar B$_{12}$ obtained from the INDO-HF
calculations, in Figs. \ref{fig:indo-molorb-icos} and \ref{Fig:indo-molorbs-planar},
respectively. Molecular orbitals presented here are close to the Fermi
level, and corresponding orbitals obtained from the first principles
DFT/B3LYP calculations are presented in Figs. \ref{fig:cage-molorb-dft}
and \ref{fig:molorb-planar-dft} of the Appendix. Orbital energies
of some of the INDO-HF orbitals are presented in Table \ref{tab:indo-orb-energies}.
Upon comparing the MOs obtained from INDO-HF calculations to those
of DFT/B3LYP calculations presented in these figures, we find perfect
agreement as to the qualitative nature of the orbitals. Regarding
the quantitative aspects, the DFT/B3LYP MOs are expectedly more diffuse
as compared to the INDO-HF ones, because they employ an extended basis
set.

Because of the inversion symmetry of icosahedral B$_{12}$, its MOs
are either symmetric (\emph{gerade}) with respect to the inversion
operation, or antisymmetric (\emph{ungerade}) with respect to it.
From Fig. \ref{fig:indo-molorb-icos} it is obvious that all the MOs
from HOMO-2 to LUMO+1 have \emph{ungerade }symmetry while LUMO+2 has
\emph{gerade} symmetry. This implies that HOMO$\rightarrow$LUMO transition
is dipole disallowed, and, therefore, the closest orbital to which
HOMO electrons can be optically excited is LUMO+2. Moreover, orbitals
LUMO and LUMO+1 are degenerate (\emph{cf}. Table \ref{tab:indo-orb-energies})
because of the symmetry. The charge density distribution of HOMO (\emph{cf}.
Fig. \ref{fig:indo-homo}) is concentrated near the two pentagons,
and vertex atoms on top/bottom of it, with negligible charge in between
the two pentagons. Thus, compared to HOMO, it can be argued that the
charge distribution of LUMO+2 (\emph{cf}. Fig. \ref{fig:indo-lumo+2})
can be obtained by transferring some charge from pentagon and top/bottom
atoms, to the region between the two pentagons.

\begin{table}
\begin{tabular}{|c|c|c|}
\hline 
Orbitals & \multicolumn{2}{c|}{Orbital Energy (eV)}\tabularnewline
\cline{2-3} 
 & Quasi-planar ($C_{3v}$) & Icosahedral ($I_{h}$)\tabularnewline
\hline
\hline 
HOMO-5 & \multicolumn{1}{c|}{-16.227} & -14.749\tabularnewline
\hline 
HOMO-4 & -15.227 & -14.719\tabularnewline
\hline 
HOMO-3 & -15.219 & -8.770\tabularnewline
\hline 
HOMO-2 & -15.219 & -8.731\tabularnewline
\hline 
HOMO-1 & -10.490 & -8.349\tabularnewline
\hline 
HOMO & -10.490 & -3.684\tabularnewline
\hline 
LUMO & 0.287 & 2.475\tabularnewline
\hline 
LUMO+1 & 0.287 & 2.475\tabularnewline
\hline 
LUMO+2 & 0.841 & 3.476\tabularnewline
\hline 
LUMO+3 & 3.833 & 3.593\tabularnewline
\hline 
LUMO+4 & 5.297 & 3.633\tabularnewline
\hline 
LUMO+5 & 5.297 & 4.028\tabularnewline
\hline
\end{tabular}

\caption{INDO canonical Hartree-Fock orbital energies of some of the orbitals
close to the Fermi level for both the quasi-planar and the cage-like
B$_{\text{12}}$clusters.}

\label{tab:indo-orb-energies}
\end{table}

Quasi-planar ($C_{3v}$) isomer of B$_{12}$ lacks inversion symmetry
so that its molecular orbitals cannot be classified as $gerade$ or
$ungerade$, a fact which is obvious from the MO plots presented in
Fig. \ref{Fig:indo-molorbs-planar}. As a result of this, the HOMO$\rightarrow$LUMO
transition is dipole allowed. However, symmetry $C_{3v}$ of the system
manifests itself in form of orbital degeneracies, with HOMO being
degenerate with HOMO-1, and LUMO with LUMO+1 (\emph{cf}. Table \ref{tab:indo-orb-energies}).
The charge distribution in HOMO-1 and HOMO is fairly delocalized with
most regions of the cluster covered, while for LUMO and LUMO+1 it
is mainly concentrated on the edge atoms, with central triangle having
negligible charge.

\subsection{Optical Absorption Spectra }

\label{sub:optics}

In this section we present the results of our INDO-CI and first-principles
TDDFT calculations of the linear optical absorption spectra of the
two isomers.

Even with a valence electron approximation the number of orbitals
involved in the INDO-CI calculation is rather large (18 occupied and
30 virtual orbitals) and can lead to very large CI expansions. We
solve this problem by freezing occupied orbitals far away from the
Fermi level. This orbital freezing is carried out in a systematic
manner, and the convergence of the results with respect to the total
number of frozen orbitals ($N_{freez}$) is carefully examined in
Appendix \ref{appa:convergence}.

For the cage, we only utilized the inversion symmetry (symmetry group
$C_{s}$), so that the orbitals and the many-electron states can be
classified into $A_{g}$ (\emph{gerade}) and $A_{u}$ (\emph{ungerade})
irreducible representations (irreps). The ground state belongs to
the $A_{g}$ irrep, while the one-photon excited states belong to
the $A_{u}$ irrep. Our final MRSDCI results for the icosahedral structure
were obtained by freezing twelve occupied orbitals ($N_{freez}=12$)
both for the $A_{g}$ and $A_{u}$ symmetry manifolds. For the quasi-planar
isomer, point-group symmetry was not used, and thus, the ground, and
the excited states, were computed in the same MRSDCI calculation.
The total number of CI configurations used in the MRSDCI calculation
is about one million in case of cage structure, and over half a million
in case of quasi-planer structure. From the sizes of these CI matrices
it is obvious that these calculations are fairly large-scale, and
that the electron-correlation effects are properly accounted for.
The detailed analysis of the convergence of our results with respect
to the number of the reference states ($N_{ref}$) used in the MRSDCI
calculations, and, therefore, the size of the CI matrix, has been
provided in Appendix. \ref{appa:convergence}

\begin{table}
\caption{Excitation energies, $E$, and many-particle wave functions of the
excited states corresponding to some of the peaks in the INDO-MRSDCI
linear absorption spectrum of icosahedral B$_{12}$ (\emph{cf}. Fig.
\ref{fig:spectrum-cage-indo}), along with the squares of their dipole
coupling ($\mu^{2}=\sum_{i}|\langle f|d_{i}|G\rangle|^{2}$) to the
ground state. $|f\rangle$ denotes the excited state in question,
$|G\rangle$, the ground state, and $d_{i}$ is the $i$-th Cartesian
component of the electric dipole operator. In the wave function, the
bracketed numbers are the CI coefficients of a given electronic configuration.
Symbols $H$/$L$ denote HOMO/LUMO orbitals. Same information about
rest of the peaks can be found in table \ref{Tab:cage-wavefunc-2}
of the Appendix \ref{appb:wavefunction}.}

\begin{tabular}{|c|c|c|l|}
\hline 
Peak	 & $E$ (eV) & $\mu^{2}$(a.u.) & Wave Function\tabularnewline
\hline
\hline 
I & 0.8745 & 0.0391 & $\vert H\rightarrow L+5\rangle$(0.5818)\tabularnewline
\hline 
 &  &  & $\vert H\rightarrow L+1;H\rightarrow L+2\rangle$(0.4295)\tabularnewline
\hline 
 &  &  & $\vert H\rightarrow L;H\rightarrow L+3\rangle$(0.3355)\tabularnewline
\hline 
 &  &  & $\vert H\rightarrow L;H\rightarrow L+6\rangle$(0.3036)\tabularnewline
\hline 
II & 1.3642 & 0.2215 & $\vert H\rightarrow L+3\rangle$(0.5821)\tabularnewline
\hline 
 &  &  & $\vert H\rightarrow L+5\rangle$(0.3354)\tabularnewline
\hline 
 &  &  & $\vert H\rightarrow L+1;H\rightarrow L+2\rangle$(0.3330)\tabularnewline
\hline 
 &  &  & $\vert H\rightarrow L;H\rightarrow L+3\rangle$(0.3152)\tabularnewline
\hline 
III & 3.4416 & 0.1486 & $\vert H\rightarrow L+2\rangle$(0.4449)\tabularnewline
\hline 
 &  &  & $\vert H\rightarrow L+1;H\rightarrow L+3\rangle$(0.3560)\tabularnewline
\hline 
 &  &  & $\vert H\rightarrow L;H\rightarrow L+2\rangle$(0.3496)\tabularnewline
\hline 
 &  &  & $\vert H\rightarrow L+4\rangle$(0.3346)\tabularnewline
\hline 
VI & 6.7030 & 0.2619 & $\vert H\rightarrow L+10\rangle$(0.3260)\tabularnewline
\hline 
 &  &  & $\vert H\rightarrow L+1;H\rightarrow L+8\rangle$(0.2449)\tabularnewline
\hline 
 &  &  & $\vert H-1\rightarrow L+2\rangle$(0.2380)\tabularnewline
\hline 
 & 6.7366 & 0.1417 & $\vert H\rightarrow L+9\rangle$(0.3778)\tabularnewline
\hline 
 &  &  & $\vert H\rightarrow L;H\rightarrow L+10\rangle$(0.2895)\tabularnewline
\hline 
 & 6.7564 & 0.3198 & $\vert H\rightarrow L+10\rangle$(0.5598)\tabularnewline
\hline 
 &  &  & $\vert H\rightarrow L;H\rightarrow L+9\rangle$(0.3254)\tabularnewline
\hline 
VIII & 7.7822 & 0.3373 & $\vert H-3\rightarrow L+6\rangle$(0.3285)\tabularnewline
\hline 
 &  &  & $\vert H-3\rightarrow L+1;H\rightarrow L+4\rangle$(0.2601)\tabularnewline
\hline 
 &  &  & $\vert H-3\rightarrow L;H\rightarrow L+3\rangle$(0.2586)\tabularnewline
\hline 
 & 7.9592 & 0.1889 & $\vert H\rightarrow L+1;H-3\rightarrow L+6\rangle$(0.2851)\tabularnewline
\hline 
 &  &  & $\vert H-2\rightarrow L+2\rangle$(0.2586)\tabularnewline
\hline 
 &  &  & $\vert H-1\rightarrow L;H\rightarrow L+1;$\tabularnewline
 &  &  & $H\text{\textrightarrow}L+4\rangle$(0.2514)\tabularnewline
\hline 
 & 8.0153 & 0.1343 & $\vert H-3\rightarrow L;H\rightarrow L+6\rangle$(0.2837)\tabularnewline
\hline 
 &  &  & $\vert H-3\rightarrow L+5\rangle$(0.2096)\tabularnewline
\hline
\end{tabular}\label{Tab:cage-wavefunc-1}
\end{table}

\begin{figure}
\vspace{2cm}

\subfloat[INDO-MRSDCI spectrum]{\includegraphics[width=8cm]{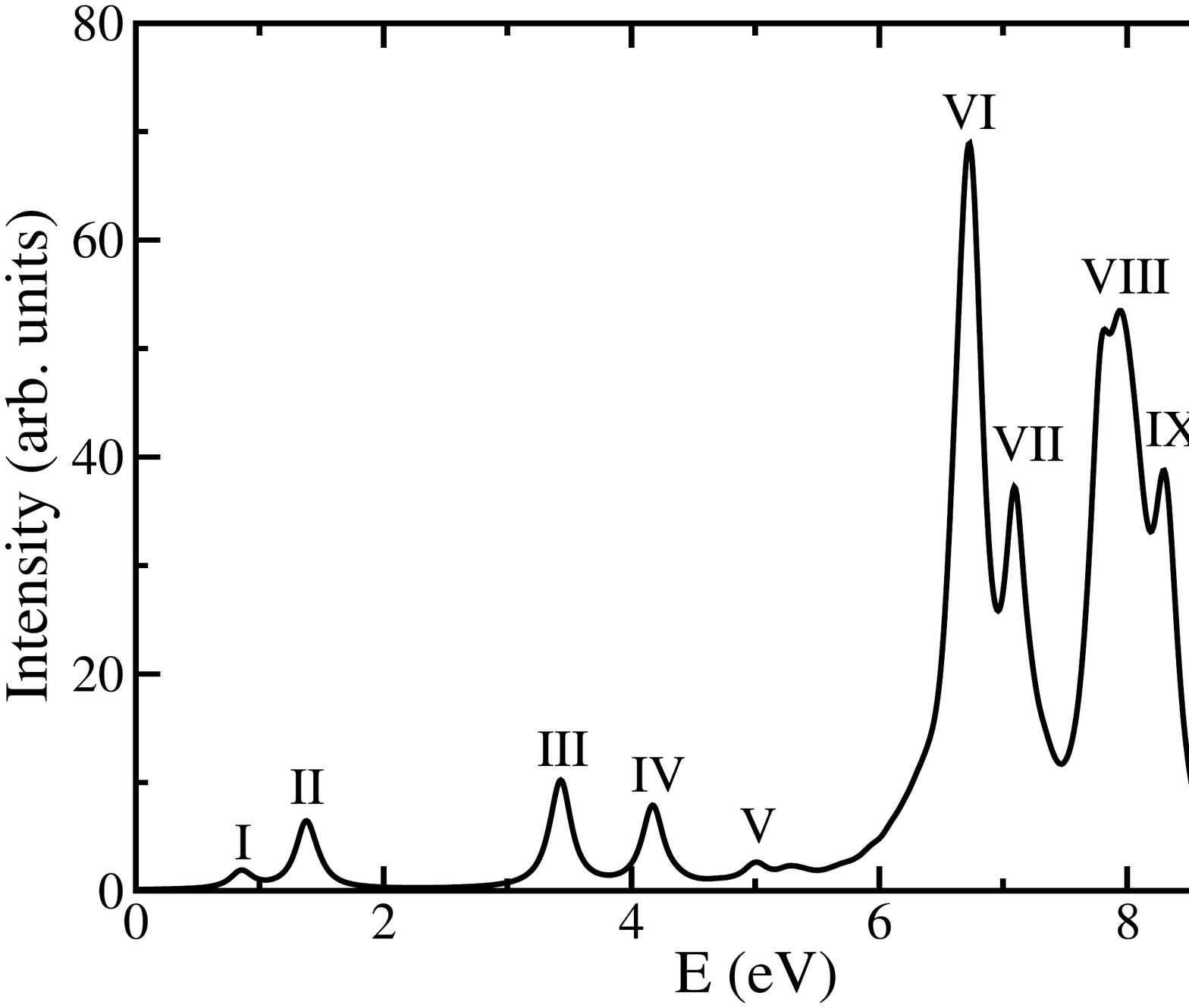}

\label{fig:spectrum-cage-indo}}

\vspace{1cm}

\subfloat[TDDFT spectrum]{\includegraphics[width=8cm]{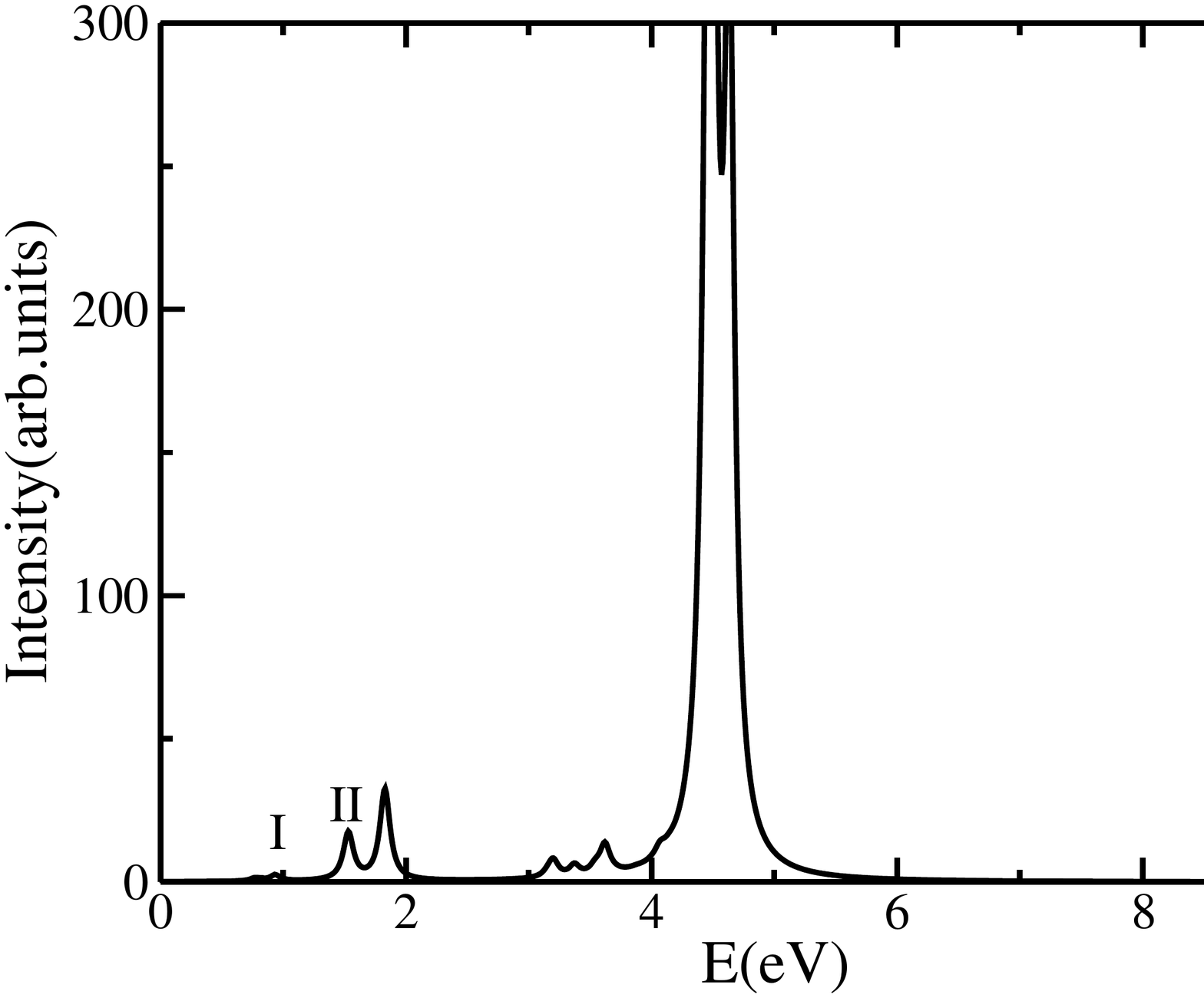}

\label{fig:spectrum-cage-tddft}}

\caption{ Linear optical absorption spectrum of icosahedral B$_{\text{12}}$
computed using: (a) the INDO-MRSDCI method , with $N_{freez}=12$
and $N_{ref}=84$, and (b) the TDDFT approach employing 6-31g(d) basis
set, and B3LYP functional. A line width of 0.1 eV was used to compute
the spectra in both the cases.}

\label{Fig:b12_cage_final} 
\end{figure}

\begin{figure}[h]
\subfloat[INDO-MRSDCI spectrum]{\includegraphics[width=8cm]{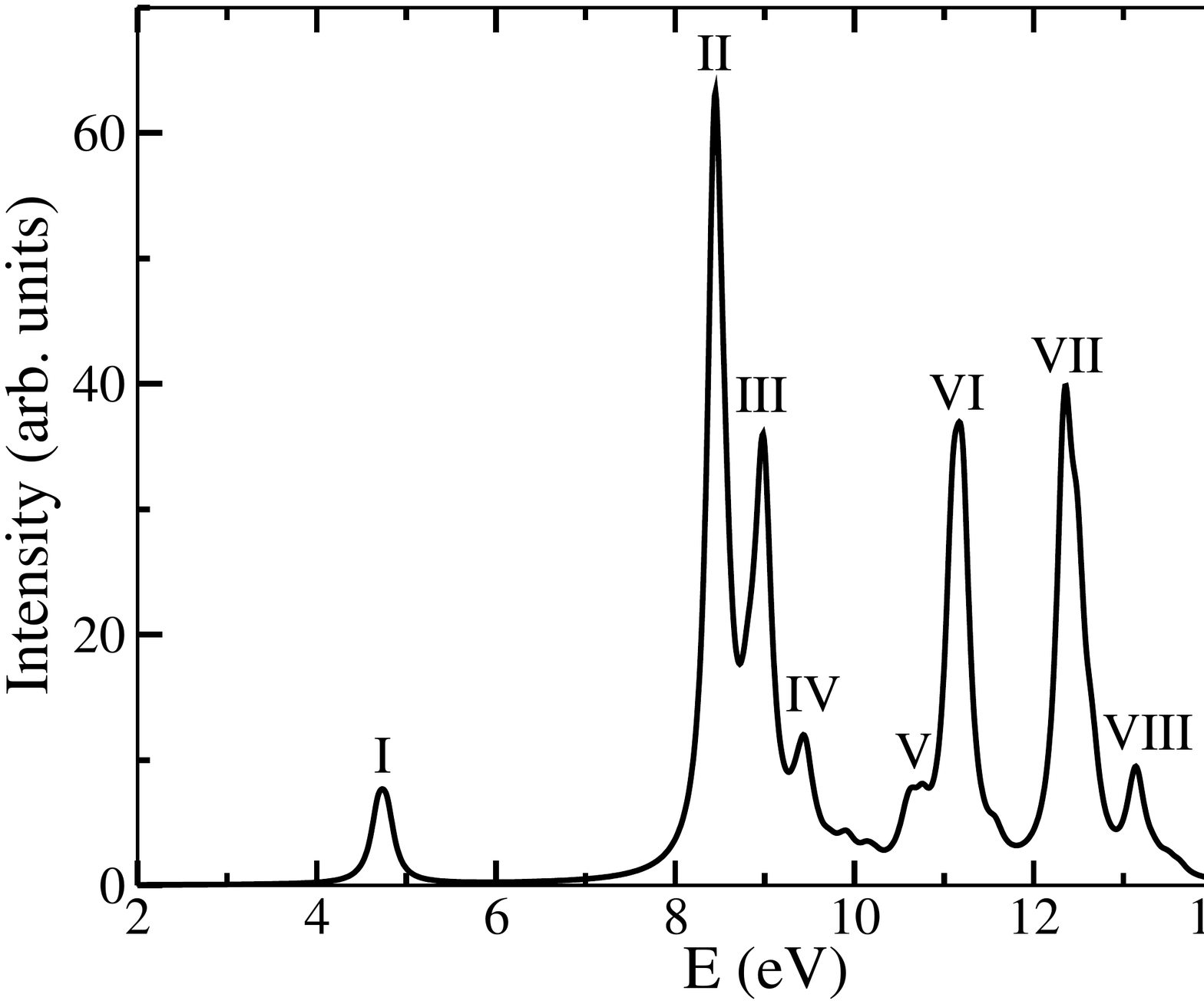}\label{fig:spectrum-indo-planar}

}

\vspace{1cm}

\subfloat[TDDFT spectrum]{\includegraphics[width=8cm]{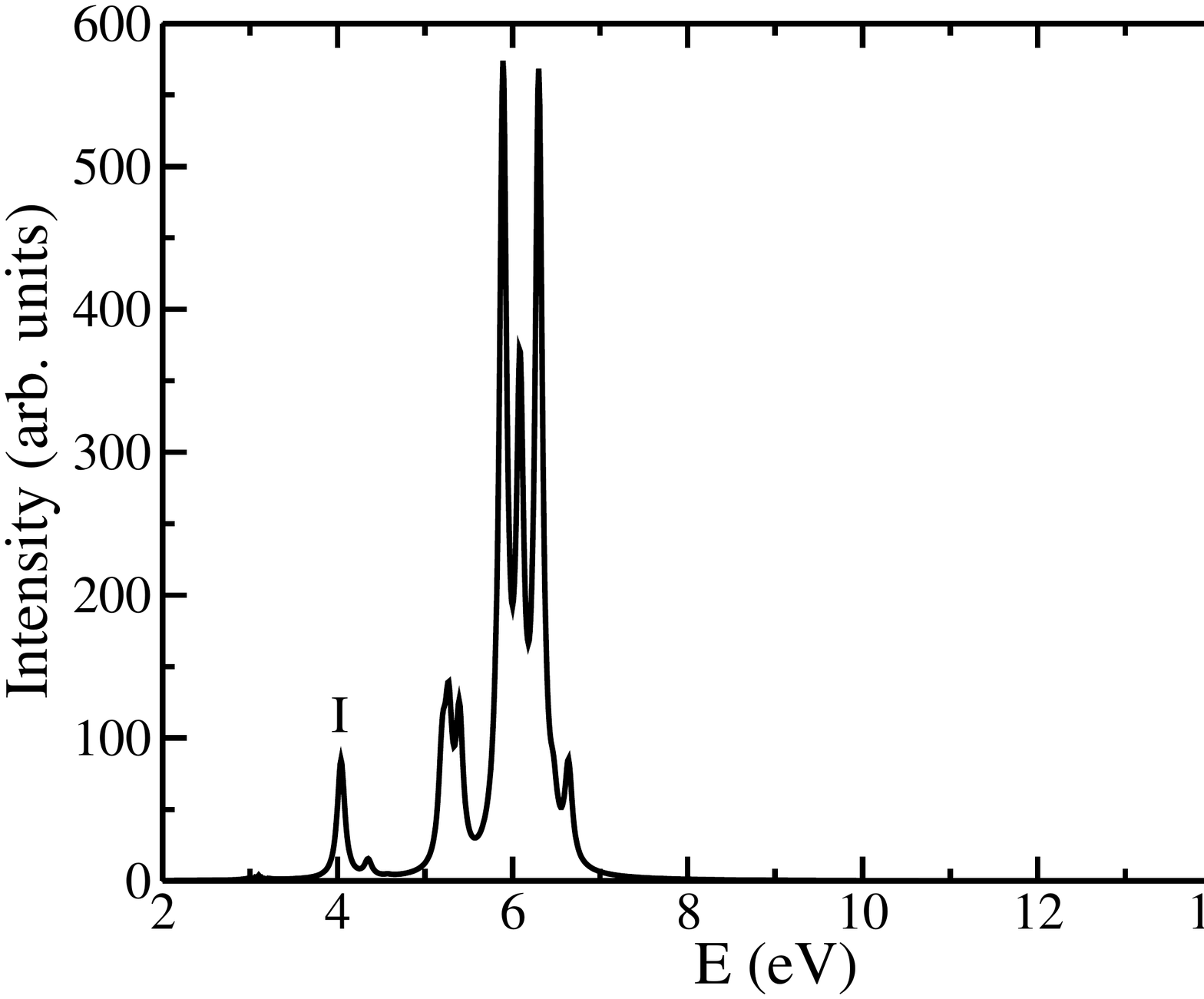}\label{fig:spectrum-tddft-planar}

}

\caption{ Linear optical absorption spectrum of quasi-planner B$_{\text{12}}$
computed using: (a) the INDO-MRSDCI method, with $N_{freez}=12$ and
$N_{ref}=31$, and (b) TDDFT method using 6-31g(d) basis set and B3LYP
functional. A line width of 0.1 eV was used to compute the spectra
in both the cases.}

\label{Fig:b12_plane_mrsd} 
\end{figure}

The linear absorption spectrum was computed using the sum-over-states
approach, with the total number of excited states ranging anywhere
from fifty to hundred. Obtaining that many excited states from large
MRSDCI calculations was a computationally expensive task, and took
several days for some calculations on our computer cluster, running
dual-CPU quad-core processors. Our INDO-MRSDCI linear absorption spectra
of the cage and planar isomers are presented in Figs. \ref{Fig:b12_cage_final}
and \ref{Fig:b12_plane_mrsd}, respectively. With the purpose of benchmarking
our INDO-CI approach, and also to study these systems from a complementary
perspective, we performed \emph{ab initio} TDDFT calculations of the
absorption spectrum of both the isomers. For the purpose we used the
same geometry as those for the INDO calculations, and used the GAUSSIAN
03 package,\cite{gaussian} coupled with the 6-31g(d) basis set, and
B3LYP gradient hybrid correlation functional. These TDDFT absorption
spectra are also presented in Figs. \ref{Fig:b12_cage_final} and
\ref{Fig:b12_plane_mrsd}.

\begin{table}
\caption{This table contains information pertinent to some of the peaks of
INDO-MRSDCI optical absorption spectrum of the quasi-planar B$_{12}$
as shown in Fig. \ref{fig:spectrum-indo-planar}. The symbols have
the same meaning as in the caption of table \ref{Tab:cage-wavefunc-1}.
Same information about rest of the peaks can be found in table \ref{Tab:plane-wave-func2}
of the Appendix \ref{appb:wavefunction}.}

\begin{tabular}{|c|c|c|c|}
\hline 
Peak	 & E(eV) & $\mu^{2}$(a.u.) & Wave function\tabularnewline
\hline
\hline 
I & 4.6844 & 0.0960 & $\vert H-1\rightarrow L+1\rangle$(0.7525)\tabularnewline
\hline 
 &  &  & $\vert H\rightarrow L\rangle$(0.5714)\tabularnewline
\hline 
 &  &  & $\vert H-1\rightarrow L;H\rightarrow L+1\rangle$(0.1271)\tabularnewline
\hline 
II & 8.4488 & 0.3599 & $\vert H\rightarrow L+3\rangle$(0.8665)\tabularnewline
\hline 
 &  &  & $\vert H-3\rightarrow L\rangle$(0.1808)\tabularnewline
\hline 
 &  &  & $\vert H\rightarrow L+4\rangle$(0.1673)\tabularnewline
\hline 
 &  &  & $\vert H-1\rightarrow L+5\rangle$(0.1361)\tabularnewline
\hline 
 &  &  & $\vert H-4\rightarrow L+1\rangle$(0.1278)\tabularnewline
\hline 
 &  &  & $\vert H\rightarrow L+1;H\rightarrow L+2\rangle$(0.3330)\tabularnewline
\hline 
III & 8.9674 & 0.1690 & $\vert H-1\rightarrow L+2\rangle$(0.4386)\tabularnewline
\hline 
 &  &  & $\vert H-1\rightarrow L+5\rangle$(0.4044)\tabularnewline
\hline 
 &  &  & $\vert H\rightarrow L+4\rangle$(0.3889)\tabularnewline
\hline 
 &  &  & $\vert H-3\rightarrow L+2\rangle$(0.3457)\tabularnewline
\hline 
 &  &  & $\vert H-2\rightarrow L+1\rangle$(0.2625)\tabularnewline
\hline 
 &  &  & $\vert H-4\rightarrow L+1\rangle$(0.2528)\tabularnewline
\hline 
 &  &  & $\vert H\rightarrow L+6\rangle$(0.2008)\tabularnewline
\hline 
IV & 9.4482 & 0.0505 & $\vert H-3\rightarrow L+1\rangle$(0.6229)\tabularnewline
\hline 
 &  &  & $\vert H-2\rightarrow L\rangle$(0.4281)\tabularnewline
\hline 
 &  &  & $\vert H-1\rightarrow L;H\rightarrow L+1\rangle$(0.2539)\tabularnewline
\hline 
 &  &  & $\vert H-1\rightarrow L+4\rangle$(0.2457)\tabularnewline
\hline 
 &  &  & $\vert H-2\rightarrow L+2\rangle$(0.2268)\tabularnewline
\hline
\end{tabular}\label{Tab:plane-wave-func1}
\end{table}

Upon comparing the INDO-MRSDCI results with the \emph{ab initio} TDDFT
ones, we conclude that the spectra computed by the two approaches
are in very good qualitative agreement with each other, with both
sets of spectra exhibiting weak absorption at lower energies, and
very intense absorptions at high energies. Next, we make a quantitative
comparison between the spectra computed by the INDO-MRSDCI and TDDFT
approaches, for both the isomers.

For the icosahedral cluster (\emph{cf}. Fig. \ref{Fig:b12_cage_final})
the two spectra exhibit the onset of optical absorption close to 0.9
eV in form of low intensity peaks, (b) next set of low intensity peaks
starts a little below 4 eV in both the spectra; in the INDO spectrum
these peaks continue beyond 4 eV, while in the TDDFT spectrum they
are all below 4 eV, and (c) high energy feature of both the spectra
are dominated by very high-intensity peaks which occur above 6 eV
in the INDO spectrum and between 4 eV and 5 eV in the TDDFT spectrum.
Thus, we have very good agreement between the spectra for the first
set of peaks around 0.9 eV and the second set of peaks at energies
close to 4 eV. For high-energy features we have a disagreement in
that the INDO theory predicts these features at energies higher than
6 eV while the \emph{ab initio} TDDFT predicts them between 4 and
5 eV. 

Upon comparing the INDO-MRSDCI and TDDFT spectra for the quasi-planar
isomer (\emph{cf}. Fig. \ref{Fig:b12_plane_mrsd}), a similar picture
emerges. In the INDO-MRSDCI spectrum the first, relatively weak, peak
occurs at around 4.7 eV followed by a set of strong peaks beyond 8
eV, with the intermediate region (4.7 eV --- 8.4 eV) exhibiting almost
negligible absorption. The TDDFT spectrum is slightly different with
the first weak peak close to 4.04 eV, followed by a slightly stronger
peak at 5.27 eV. Really intense peaks in the TDDFT spectrum occur
in the energy range between 5.9 eV and 7 eV. Therefore, the quantitative
comparison between the INDO-MRSDCI and TDDFT spectra for the quasi-planar
isomer is quite similar to that of the icosahedral isomer. Agreement
is reasonable for the lower energy peaks, but for the higher peaks,
TDDFT spectrum is significantly redshifted as compared to the INDO-MRSDCI
spectrum. Qualitatively, however, both sets of spectra are in very
good agreement with each other.

\begin{figure}
\includegraphics[width=8cm]{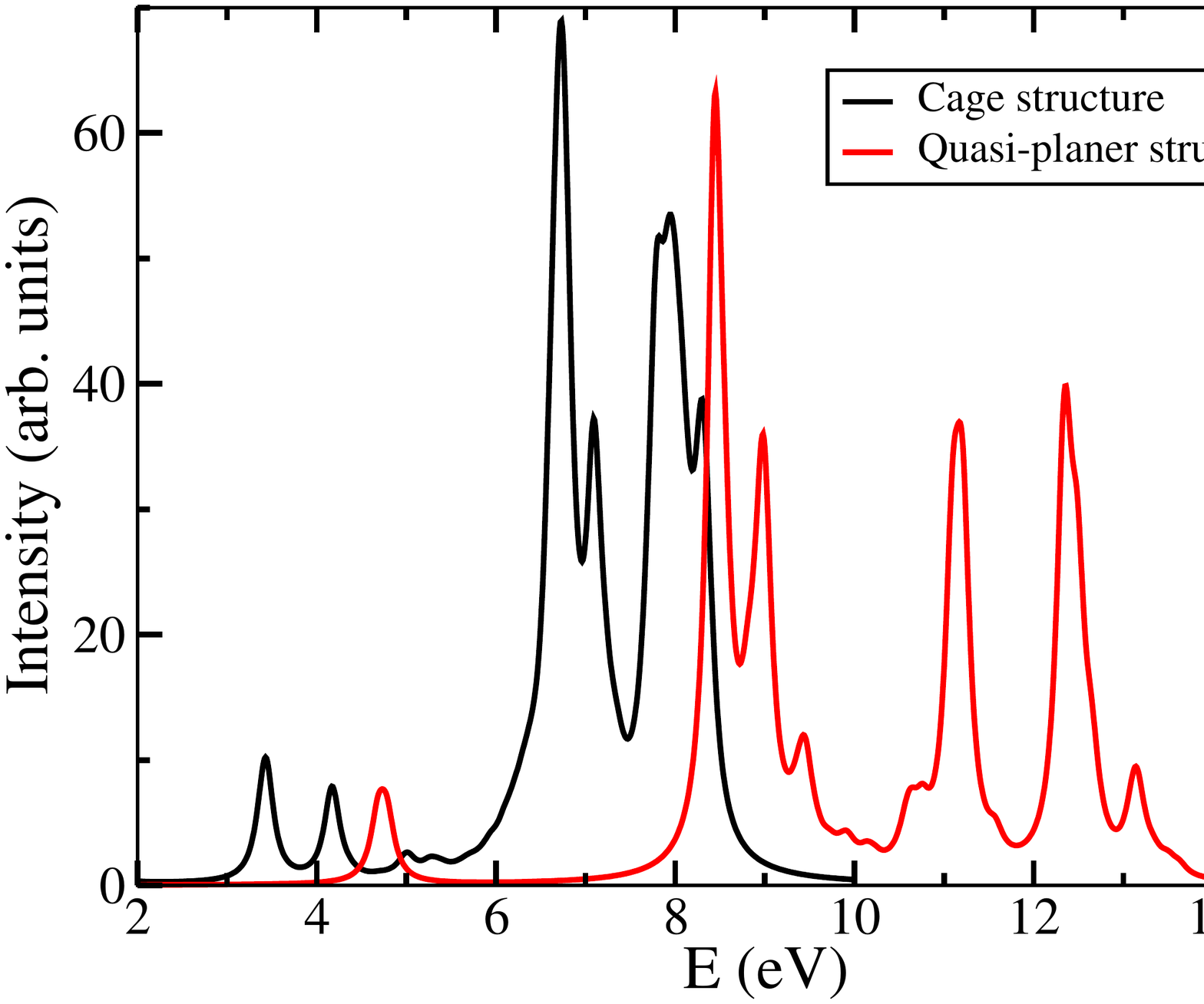}

\caption{(color online) Comparison of the linear optical spectra of icosahedral
(black) and quasi-planar (red) structures of $B_{12}$cluster.}

\label{Fig:compare}
\end{figure}

Next we present a detailed comparative analysis of our INDO-MRSDCI
results for the cage and the quasi-planar structures. To facilitate
this comparison visually, we also present a combined plot of the two
spectra in Fig. \ref{Fig:compare}. A cursory look at Fig. \ref{Fig:compare}
reveals that lower energy regions of the absorption spectra consist
of relatively lower intensity peaks while the higher energy spectra
in both the isomers exhibit intense absorption. A comparison of the
the spectra of the two isomers reveals that: (a) the absorption spectrum
of the icosahedral isomer is significantly red-shifted as compared
to the planar one, and (b) the peak intensities in both the spectra
are of the same order of magnitude. The main distinguishing feature
of the absorption spectra of the two isomers, as depicted by our calculations,
is that in the icosahedral B$_{12}$ the optical absorption (\emph{cf}.
Fig. \ref{Fig:b12_cage_final}) begins at rather low energies with
the first peak (peak I) located at 0.86 eV, with two more peaks (peaks
II and III) located below 4 eV, while in the quasi-planar isomer (Fig.
\ref{Fig:b12_plane_mrsd}) the first absorption absorption feature
(peak I) is located significantly higher at 4.73 eV. This INDO-MRSDCI
result is also confirmed in the TDDFT calculations, which also predict
the absorption to commence at much lower energies in the icosahedral
isomer, as compared to the quasi-planar one (\emph{cf}. Figs. \ref{fig:spectrum-cage-tddft}
and \ref{fig:spectrum-tddft-planar}). This prediction of our calculations
can be tested in future experiments on boron clusters, and if confirmed,
can be used for distinguishing the icosahedral clusters from the quasi-planar
ones in optical absorption experiments.

As far as the higher energy features are concerned, in the icosahedral
cluster (Fig. \ref{Fig:b12_cage_final}), a series of high intensity
features (peaks V, VI, VII, and VIII) start around 6.7 eV and continue
past 8 eV. In the planar isomer (Fig. \ref{Fig:b12_plane_mrsd}) on
the other hand the higher energy absorptions (peaks II through VII)
start beyond 8 eV and continue to much higher energies beyond 12 eV.

The many-particle wave functions of the excited states corresponding
to some of the peaks in the spectra, along with the squares of their
transition dipoles, are presented in Tables \ref{Tab:cage-wavefunc-1}
and \ref{Tab:plane-wave-func1}. The same information about the rest
of the peaks can be examined in tables \ref{Tab:cage-wavefunc-2}
and \ref{Tab:plane-wave-func2} of Appendix \ref{appb:wavefunction}.
One distinguishing feature of the optical absorption in cage, compared
to that in the quasi-planar isomer is that HOMO ($H$) to LUMO ($L$)
transition is dipole forbidden in the cage because, as discussed above,
$H$ and $L$ orbitals have the same inversion symmetry of $u$, and
among the unoccupied orbitals, $L+2$ is the lowest-lying with the
opposite symmetry of $g$, thereby making $H\rightarrow L+2$ as the
lowest allowed single-particle optical transition for the cage. For
the quasi-planar B$_{12}$ on the other hand, the single-particle
transition $H\rightarrow L$ is optically allowed because the system
does not have inversion symmetry.

Before further discussions of our results it is useful to remember
that in the crystalline form boron is a semiconductor, with an indirect
band gap close to 1.53 eV for $\beta$-rhombohedral boron,\cite{boron-band-gap}
whose crystal structure is also based on B$_{12}$ icosahedron. Semiconductors
exhibit single-particle inter-band optical absorption at lower energies,
and collective plasmon absorption at high energies. Therefore, it
is of interest to understand the optical absorption spectra of B$_{12}$
clusters based on these pictures and to ascertain whether absorption
exhibits single-particle behavior or the collective one. Indeed for
the case of metallic clusters such as those of alkali metals the issue,
whether the optical excitations are similar to those in a molecule,
or they are related to the plasmons in the bulk, has been extensively
examined.\cite{heer-rmp} In metal cluster physics plasmon excitations
are identified by appearance of continuous absorption pattern. Moreover,
Koutecký and coworkers\cite{koutecky} have devised a scheme according
to which if the many particle wave function of a given excited state
is dominated by one singly-excited configuration it is classified
as a normal inter-band absorption. %

On the other hand if the excited state wave function is a linear combination
of several configurations with similar weights, it is called a collective
(plasmon like) excitation.\cite{koutecky} Close examination of the
many-particle wave functions of both the isomers presented in Tables
\ref{Tab:cage-wavefunc-1}, \ref{Tab:plane-wave-func1}, \ref{Tab:cage-wavefunc-2},
and \ref{Tab:plane-wave-func2} reveals that none of the excited states
participating in the optical absorption of B$_{12}$ clusters are
dominated by single configurations. All these excited states have
prominent multi-reference character, and are linear combinations of
several singly- and doubly-excited configurations with significant
coefficients. For example, peaks I of both the isomers are a mixture
of singly and doubly excited configurations, as are all the other
peaks. This, as per the criterion outlined in the work of Koutecký
and coworkers,\cite{koutecky} suggests plasmon like behavior. Moreover,
the absorption spectrum appears to be fairly continuous starting with
low-energy and low-intensity absorptions, followed by high-energy
high-intensity absorptions before tapering off.

\section{Conclusions and Future Directions}

\label{sec:conclusion}

In conclusion, we have presented a theoretical study of linear optical
absorption in two B$_{12}$ isomers, namely icosahedral and quasi-planar
clusters. The wave-function-based calculations were performed using
the semi-empirical INDO model,\cite{INDO} and electron correlation
effects were taken into account by means of large-scale MRSDCI computations.
To obtain an alternative perspective on the optical absorption, TDDFT
method based \emph{ab initio} calculations of the spectrum were also
performed. A comparison of the INDO-MRSDCI and TDDFT calculations
revealed that the two methods lead to spectra which are qualitatively
very similar. On the quantitative front, the high-energy features
of spectra were found to be red-shifted in the TDDFT approach as compared
to the INDO-MRSDCI approach. Which of these results is closer to reality
can only be decided by the experiments, which, we hope will be performed
in the future. Another aspect of these spectra is that the optical
absorption in the icosahedral clusters begins at much lower energies
as compared to the planar one, a fact which can be used in the optical
detection of these clusters. The high-intensity absorption in the
cluster takes place at higher energies, which our calculations suggest,
are plasmonic in nature. Results of these calculations, which to the
best of our knowledge have not been performed earlier, can be tested
in the future experiments. It will also be interesting to investigate
the nature of triplet excited states in boron based clusters along
with their nonlinear optical properties. Calculations along those
directions are in progress in our group, and results will be presented
in future publications.

\appendix

\section{Moleular orbitals of B$_{\text{12}}$ isomers obtained from \emph{ab
initio} Calculations}

\label{sec:app-molorb}

\begin{figure}
\subfloat[HOMO-2]{\includegraphics[angle=90,scale=0.15]{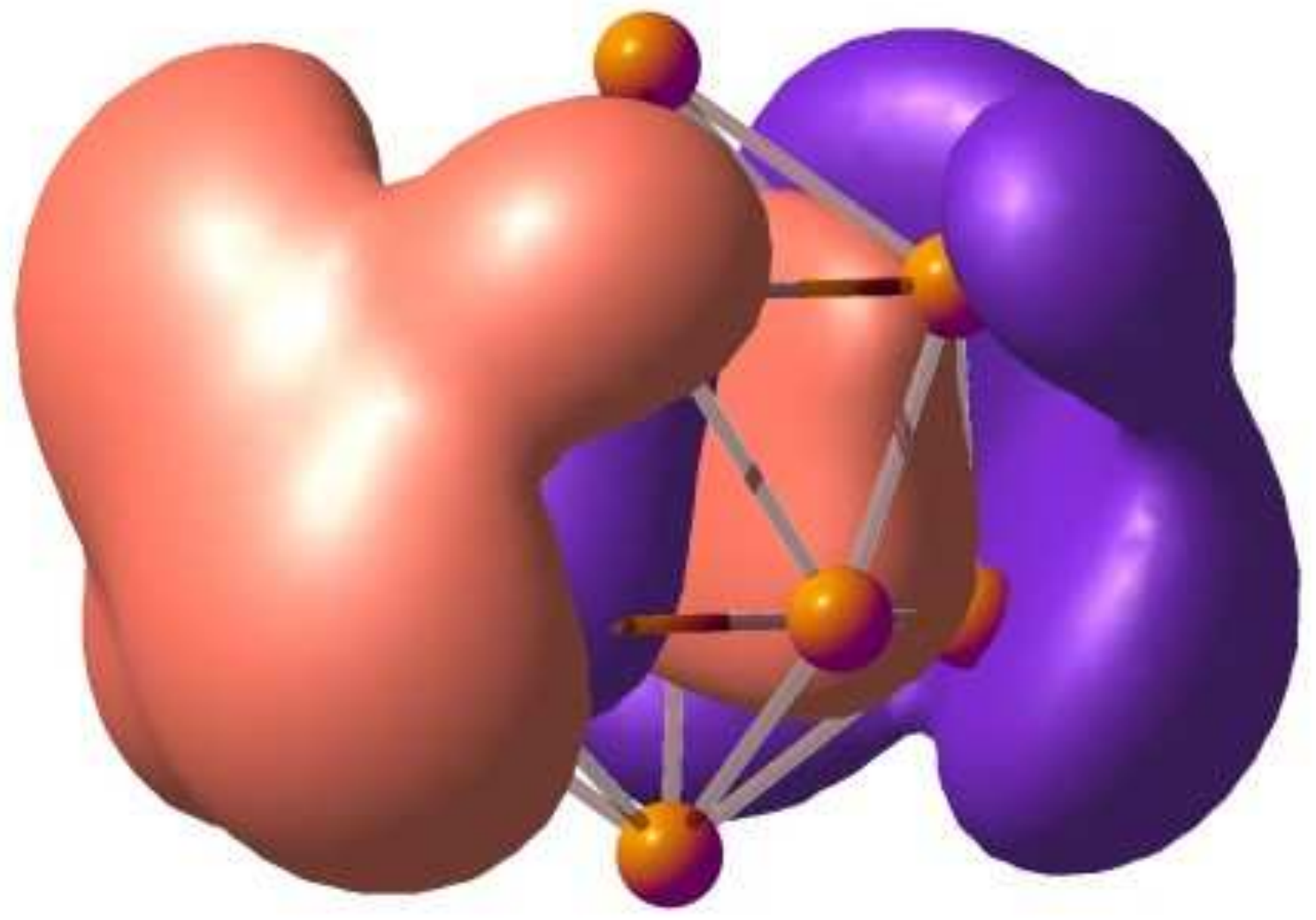}

}\subfloat[HOMO-1]{\includegraphics[scale=0.15]{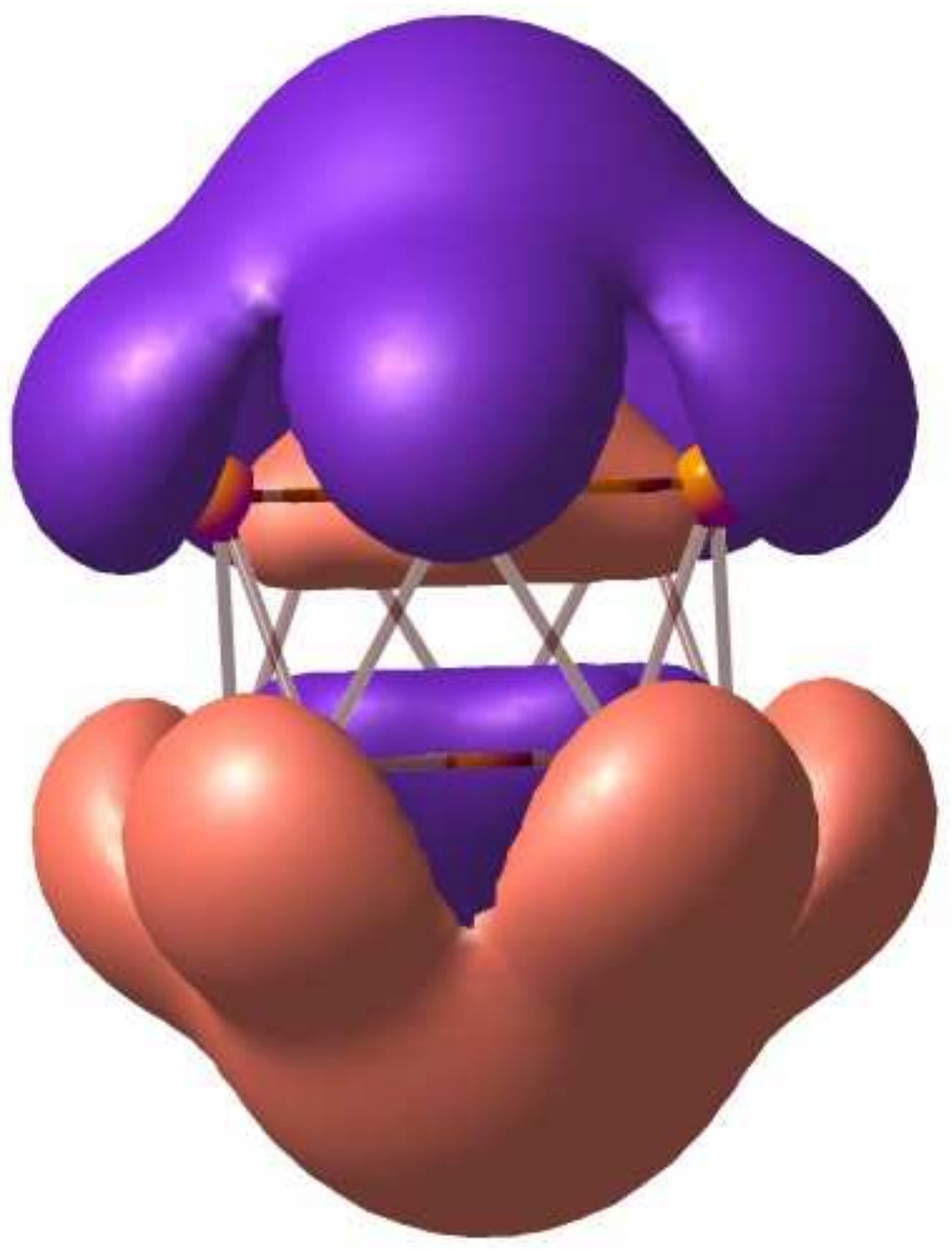}

}

\subfloat[HOMO]{\includegraphics[scale=0.15]{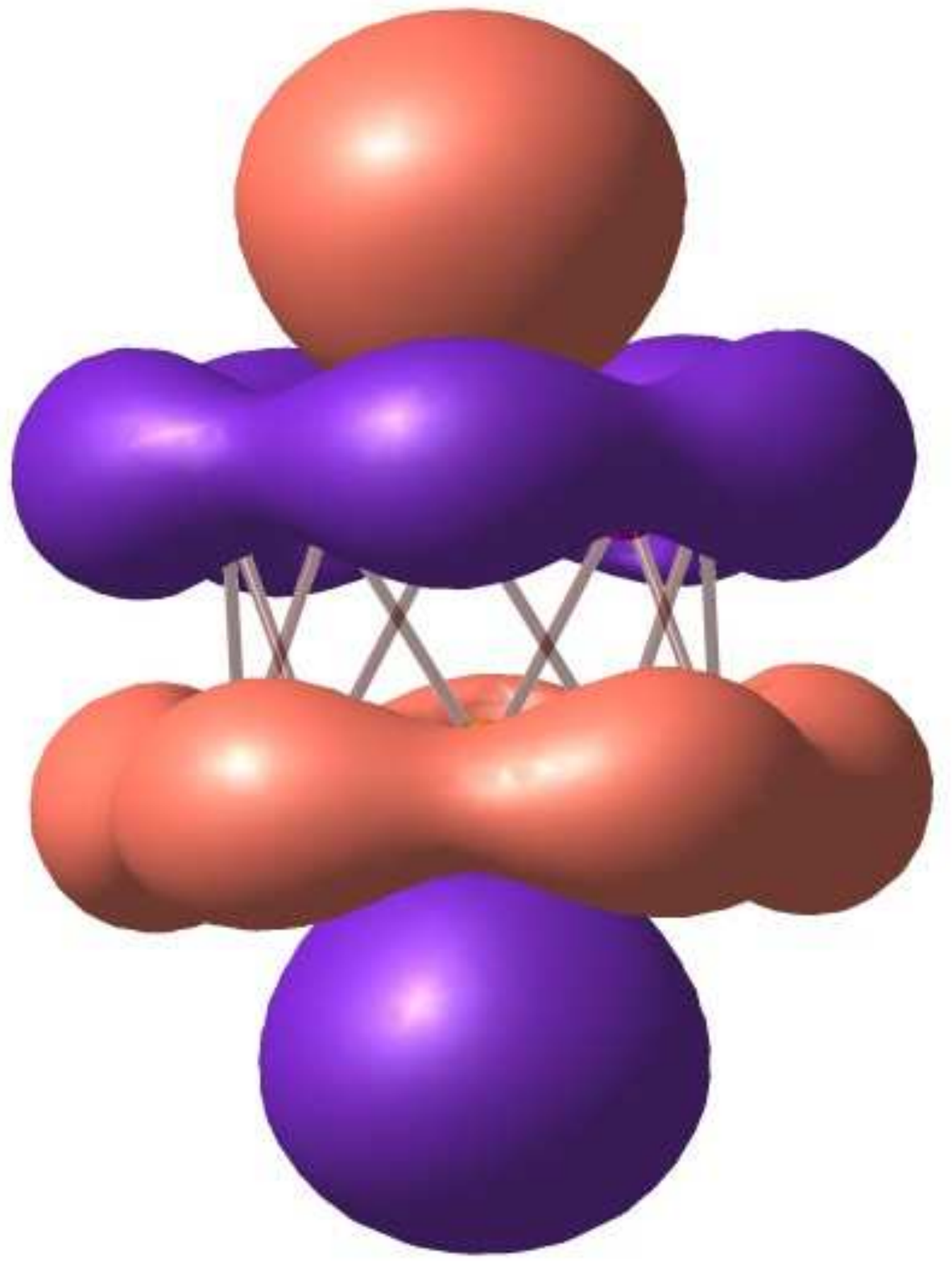}

}\subfloat[LUMO]{\includegraphics[scale=0.15]{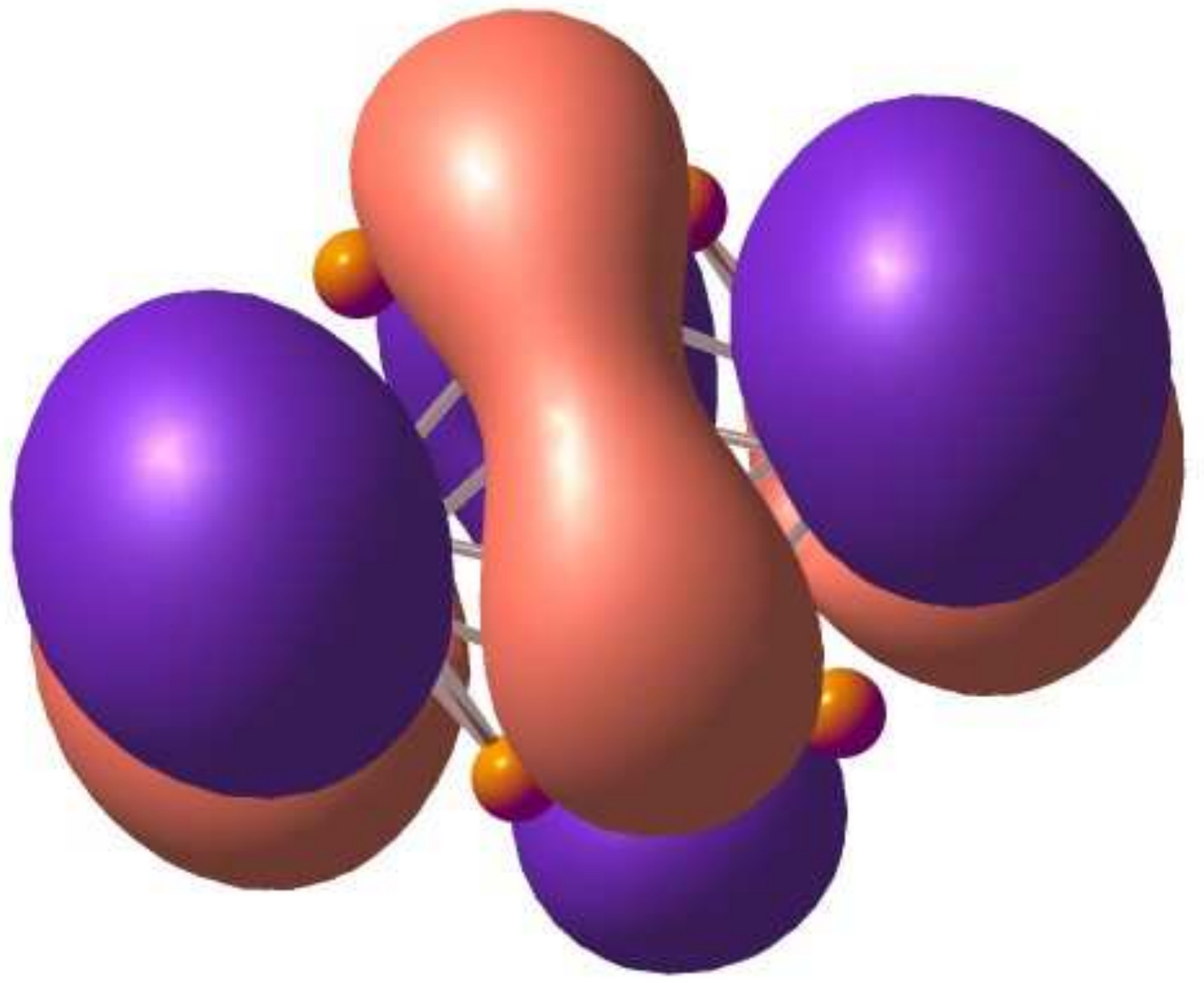}

}

\subfloat[LUMO+1]{\includegraphics[scale=0.15]{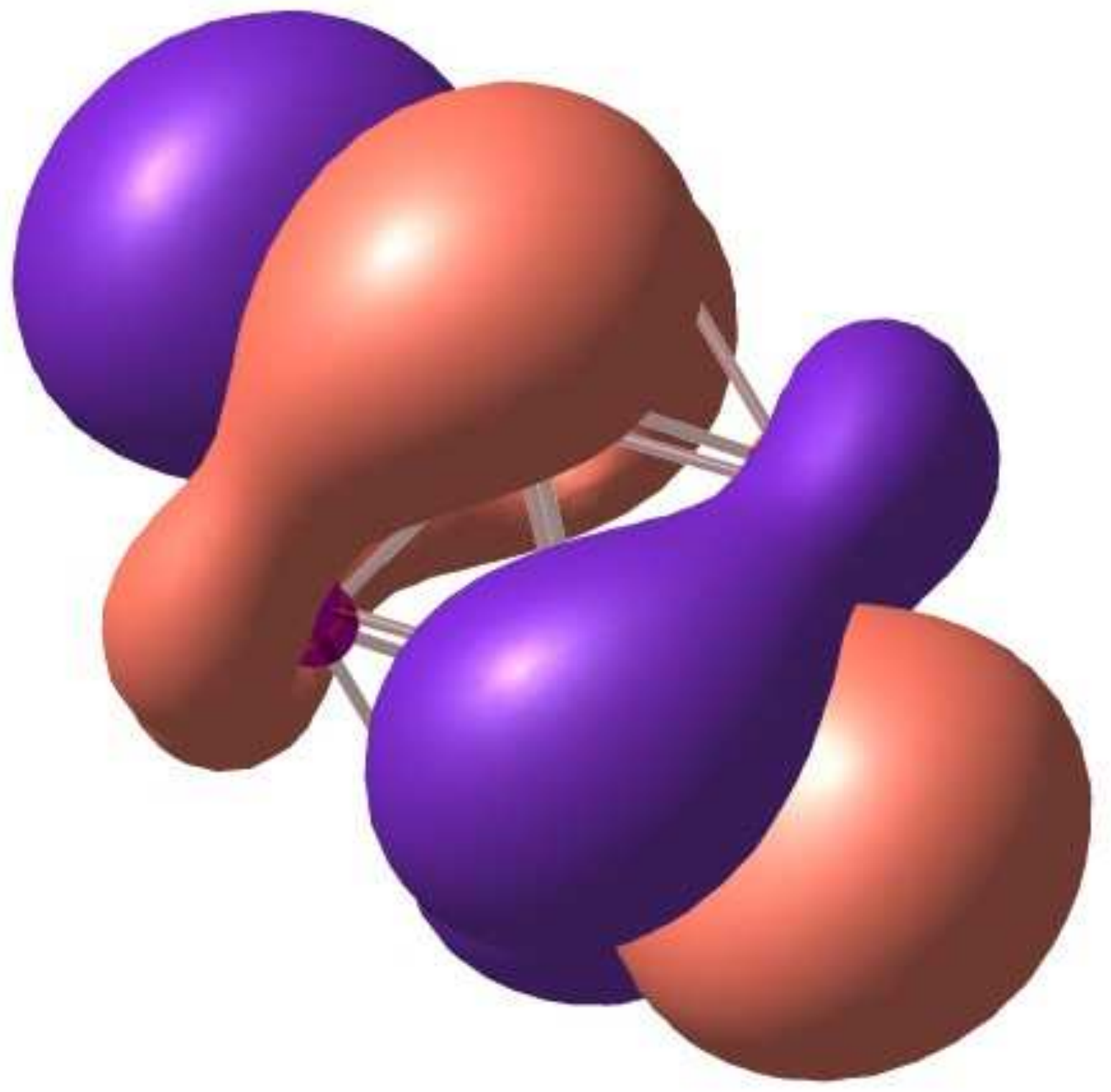}

}\subfloat[LUMO+2]{\includegraphics[scale=0.15]{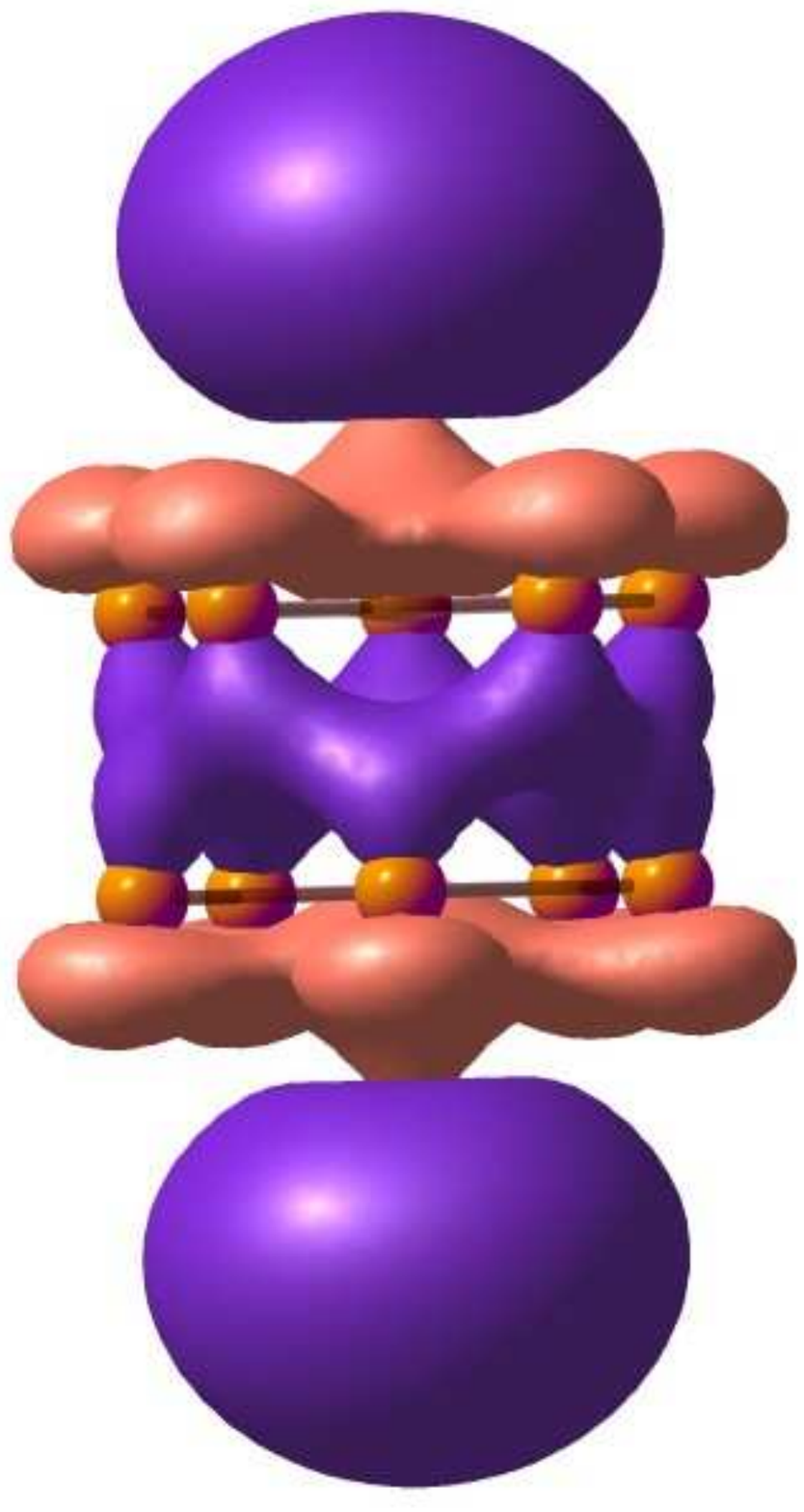}

}

\caption{(color online) Some molecular orbitals (iso plots) of icosahedral
B$_{\text{12}}$, obtained from the first principles DFT/B3LYP calculations.}

\label{fig:cage-molorb-dft}
\end{figure}

\begin{figure}
\subfloat[HOMO-1]{\includegraphics[scale=0.15]{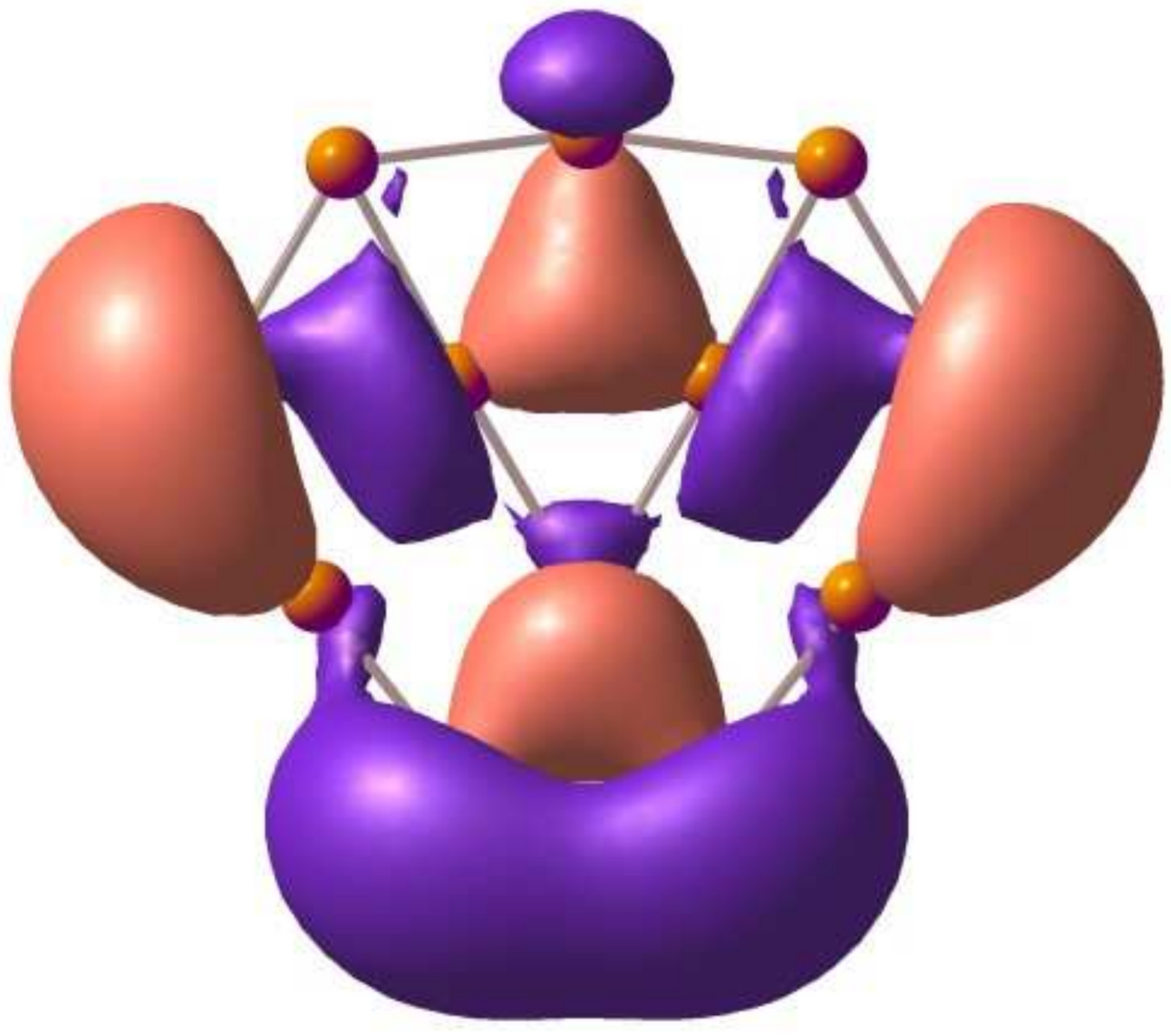}

}\subfloat[HOMO]{\includegraphics[scale=0.15]{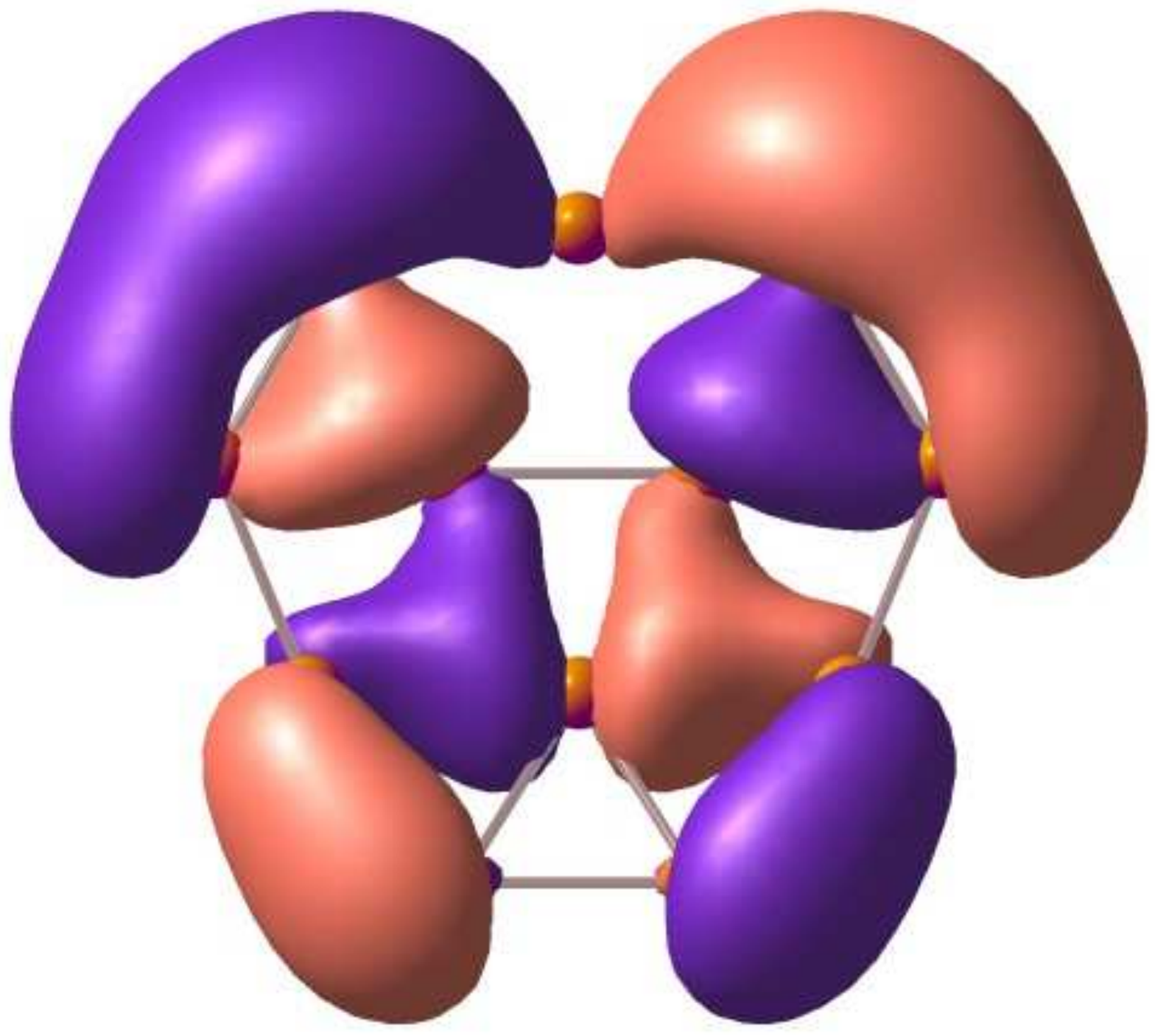}

}\subfloat[LUMO]{\includegraphics[scale=0.15]{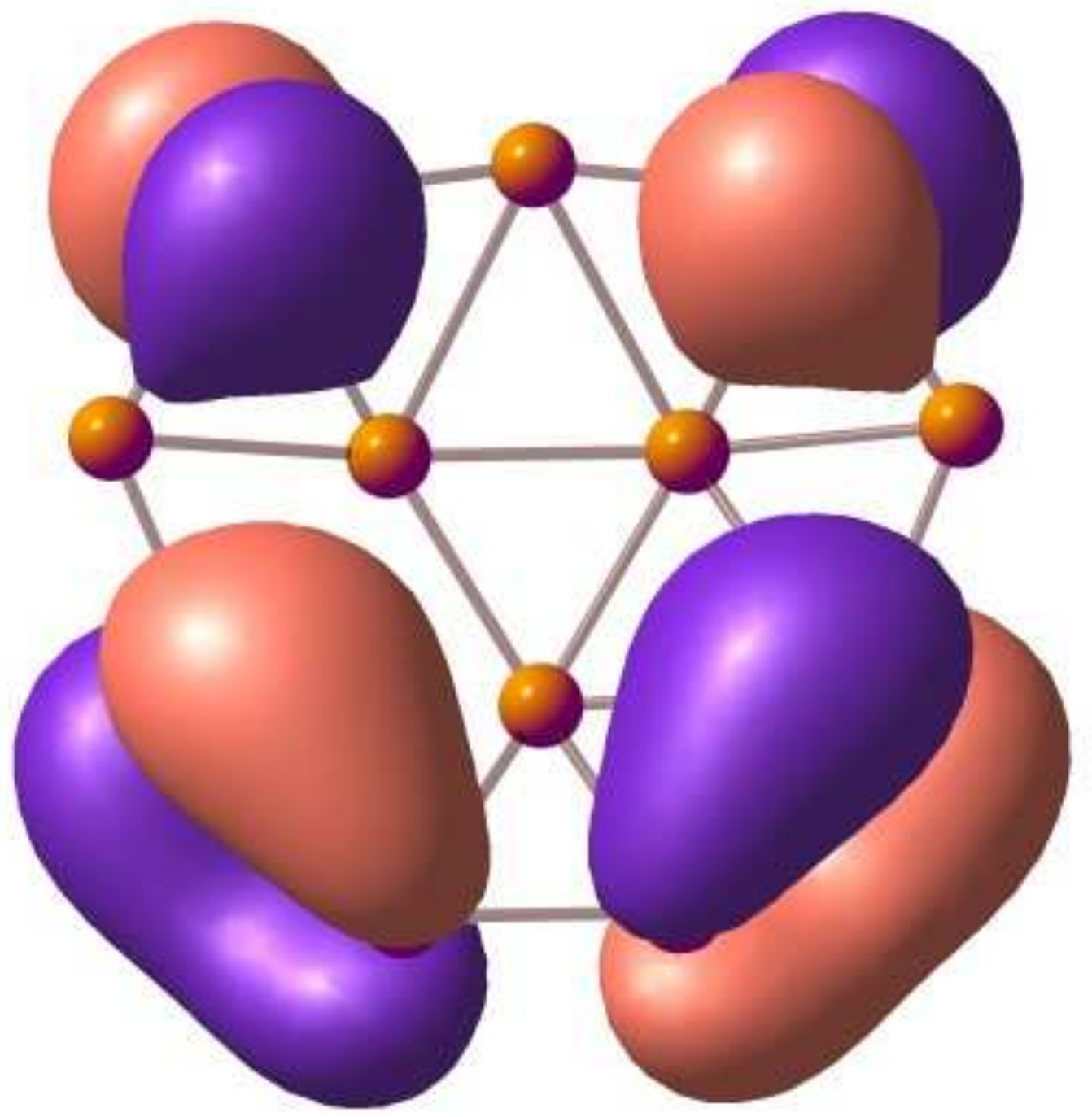}

}

\subfloat[LUMO+1]{\includegraphics[scale=0.15]{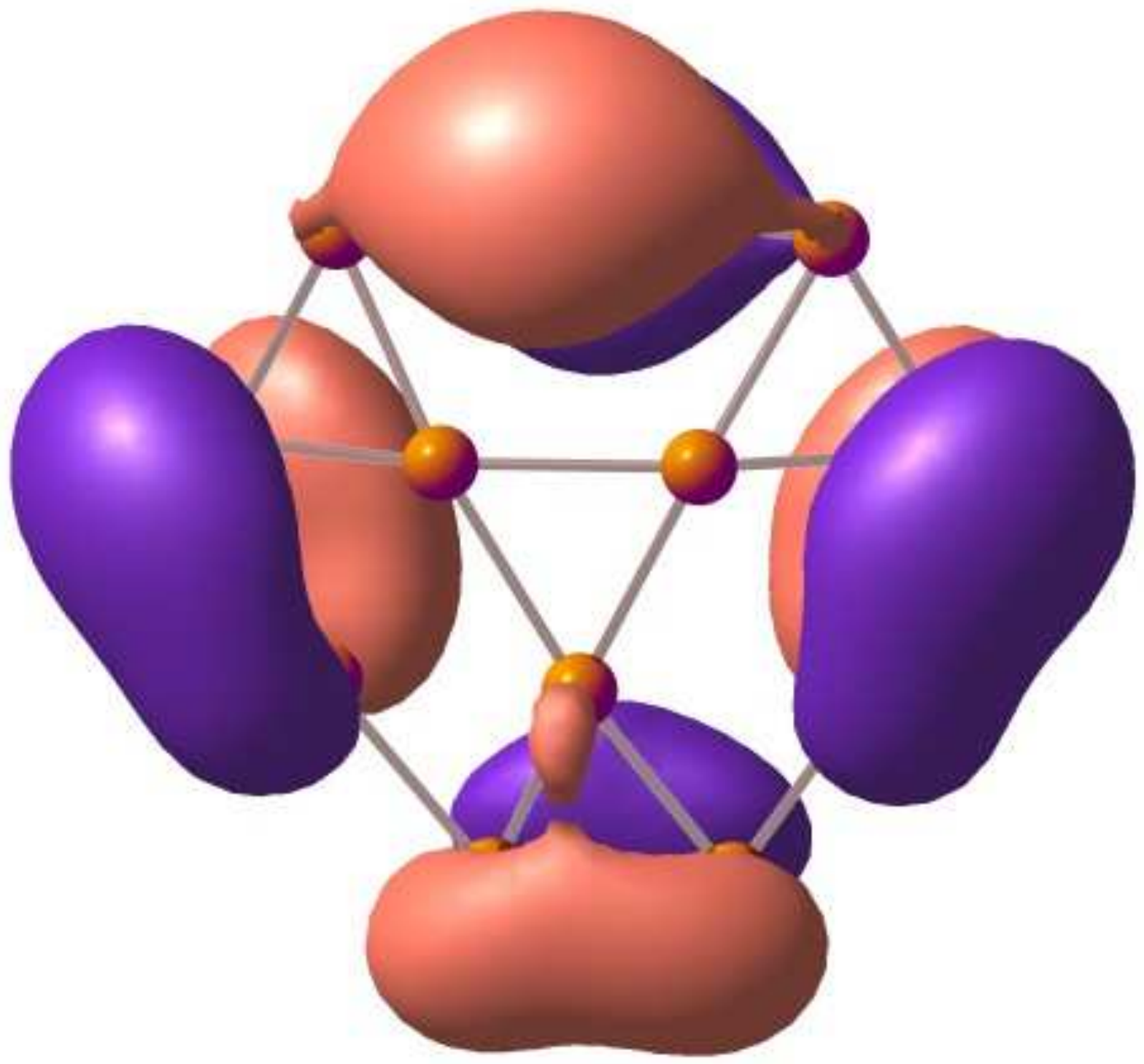}

}

\caption{(color online) Some molecular orbitals (iso plots) of quasi-planar
B$_{\text{12}}$, obtained from the first principles DFT/B3LYP calculations.}

\label{fig:molorb-planar-dft}
\end{figure}

In this section we present some of the molecular orbitals of icosahedral
and quasi-planar isomers of B$_{12}$, obtained from the first-principles
DFT calculations employing B3LYP functional, and the 6-31g(d) gaussian
basis set. The calculations were performed using the Gaussian03 program,\cite{gaussian}
and the orbitals presented are the ones close to the Fermi level.
In Fig. \ref{fig:cage-molorb-dft} orbitals of icosahedral isomer
are plotted, while in Fig. \ref{fig:molorb-planar-dft} those of the
quasi-planar isomer are presented. These orbitals agree quite well
with the corresponding orbitals obtained from the INDO-HF calculations,
presented in Figs. \ref{fig:indo-molorb-icos} and \ref{Fig:indo-molorbs-planar}
of the main text. 

\section{Convergence Issues}

\label{appa:convergence}

\begin{figure}
\includegraphics[width=8cm]{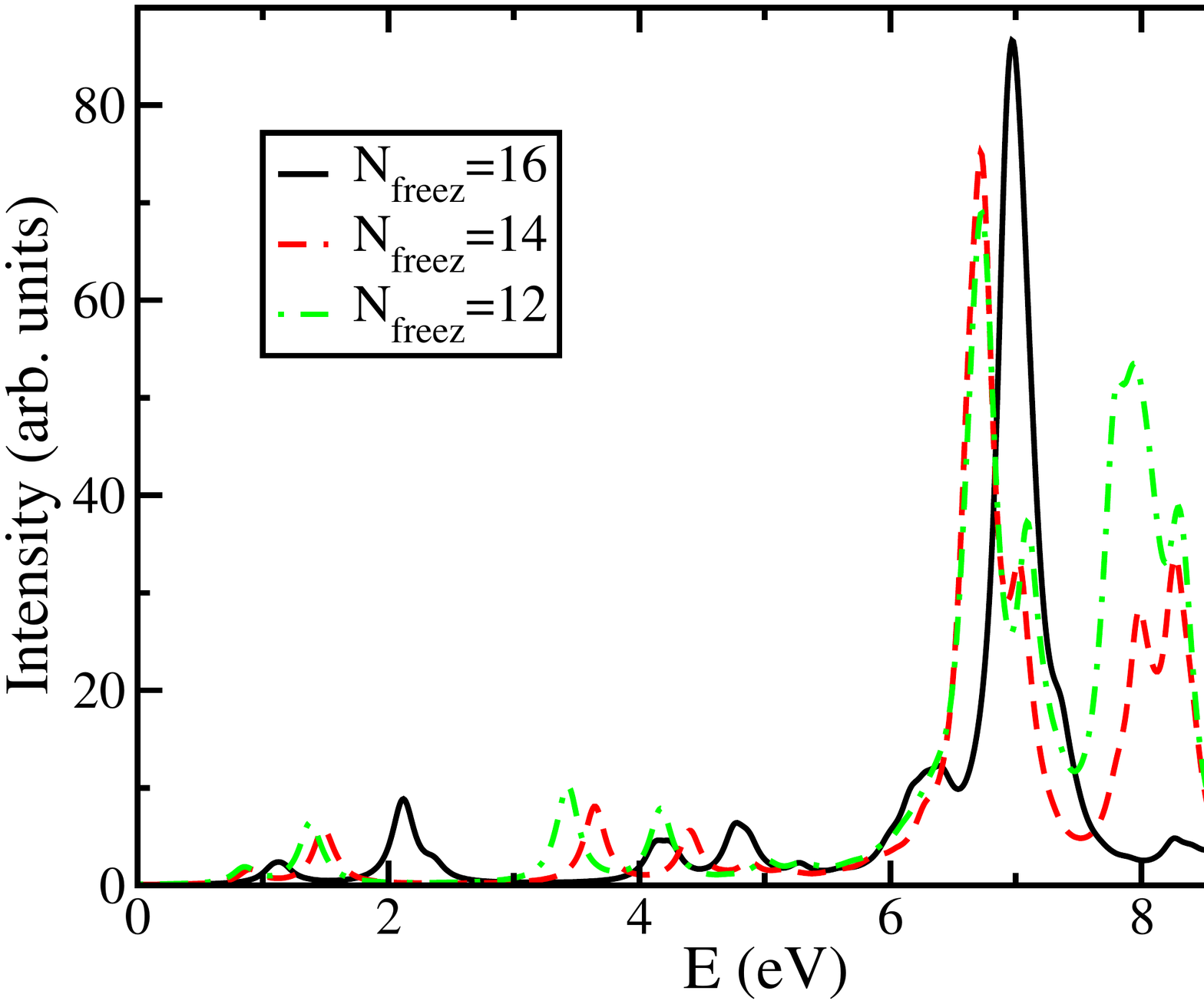}

\caption{(color online) Convergence of the linear absorption spectrum of icosahedral
B$_{\text{12}}$ computed using the INDO-MRSDCI method for decreasing
number of frozen occupied orbitals. A line width of 0.1 eV was used
to compute the spectra.}

\label{Fig:cage_converg}
\end{figure}

\begin{figure}[h]
\vspace*{1cm}
\includegraphics[width=8cm]{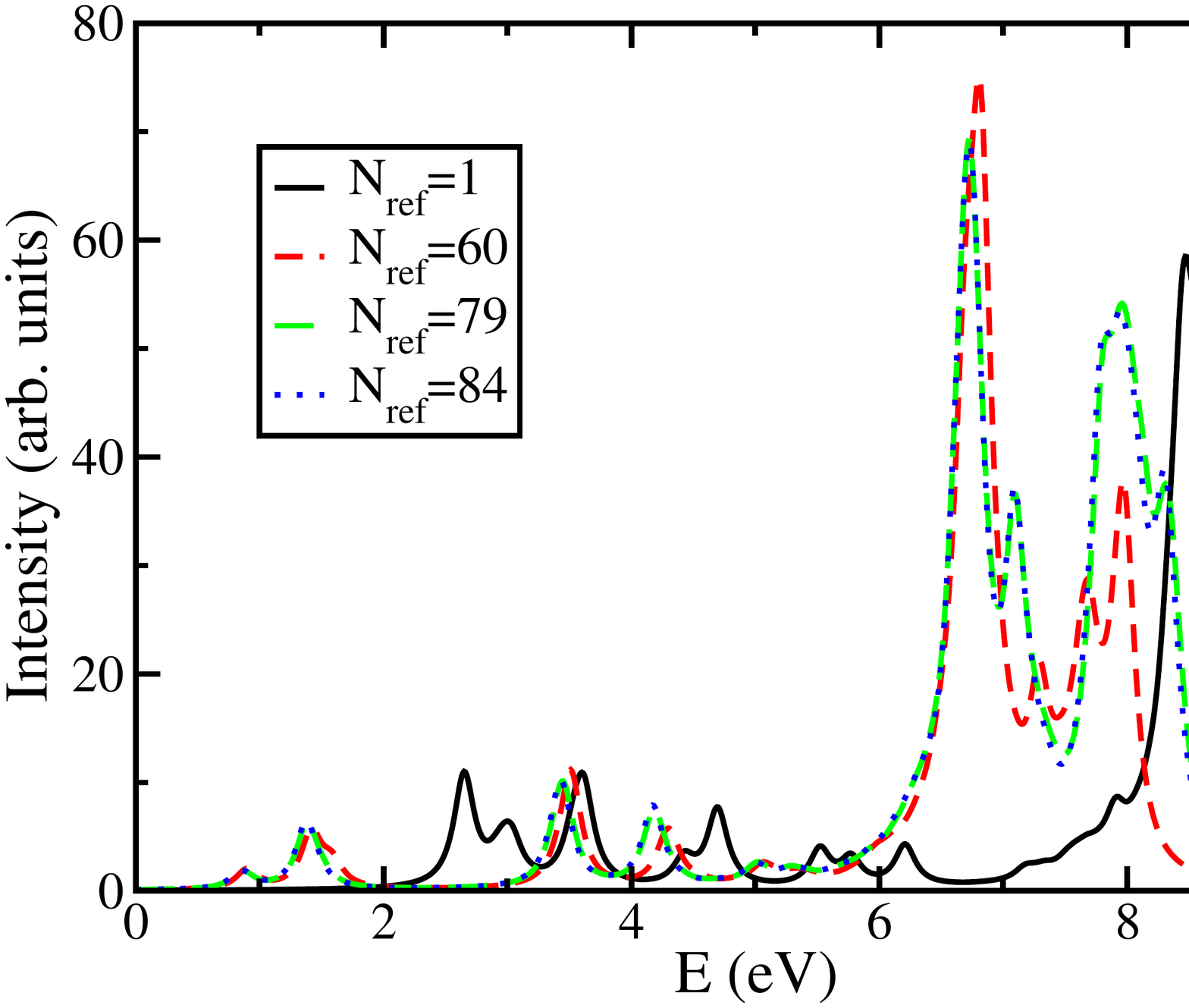}

\caption{(color online) Convergence of the linear absorption spectrum of icosahedral
B$_{\text{12}}$ with increasing number of reference configurations
($N_{ref}$) in the MRSDCI wave function with $N_{freez}=12$. A line
width of 0.1 eV was used to compute the spectra.}

\label{Fig:cage_converge_mrsd} 
\end{figure}

In this section we discuss the convergence of our MRSDCI calculations
with respect to: (a) the number of frozen orbitals $N_{freez}$, and
(b) the number of reference configurations $N_{ref}$, with respect
to which the singly- and doubly-excited configurations spanning the
CI space are generated. For a given value of $N_{freez}$, successively
larger MRSDCI calculations (\emph{i.e.} with the increasing values
of $N_{ref}$) were performed, until no significant changes in the
results were observed both with respect to, decreasing $N_{freez}$,
and increasing $N_{ref}$. In Fig. \ref{Fig:cage_converg} we present
the results of the best (largest $N_{ref}$) MRSDCI calculations on
the icosahedral B$_{12}$ with $N_{freez}=16,\:14,$ and 12, and $N_{ref}=83$,
92, and 84, respectively. In these calculations the sizes of the MRSDCI
matrices ($N_{total}$) ranged from 20935 for $N_{freez}=16$ to 940945
for $N_{freez}=12$. From the figure it is obvious that there are
no significant qualitative or quantitative changes in the absorption
spectrum as $N_{freez}$ is decreased from fourteen to twelve. Similarly,
Fig. \ref{Fig:cage_converge_mrsd} demonstrates the convergence of
the absorption spectrum with respect to increasing $N_{ref}$ (and
hence $N_{tot}$), with values $N_{ref}=1,\:60,\:79,$ and 84. Thus,
we can conclude that our results for the cage B$_{12}$ are well converged
both with respect to $N_{freez}$ as well as $N_{ref}$.

Similar checks of convergence were also performed for the quasi-planar
isomer of B$_{12}$, figures corresponding to which are not presented
here for the reasons of brevity.

\section{Excited State Wave Functions, Energies, and Transition Moments}

\label{appb:wavefunction}

In the following two tables we present the excitation energies, the
many-particle wave functions, and the transition dipole moments with
respect to the ground state, of the excited states corresponding to
those peaks in the INDO-MRSDCI linear absorption spectra of the two
isomers of $B_{12}$, which were not included in the tables \ref{Tab:cage-wavefunc-1}
and \ref{Tab:plane-wave-func1} of Sec. \ref{sec:results}.

\begin{table}
\caption{Excitation energies, $E$, and many-particle wave functions of the
excited states corresponding to some of the peaks in the INDO-MRSDCI
linear absorption spectrum of icosahedral B$_{12}$ (\emph{cf}. Fig.
\ref{fig:spectrum-cage-indo}), along with the squares of their dipole
coupling ($\mu^{2}=\sum_{i}|\langle f|d_{i}|G\rangle|^{2}$) to the
ground state. $|f\rangle$ denotes the excited state in question,
$|G\rangle$, the ground state, and $d_{i}$ is the $i$-th Cartesian
component of the electric dipole operator. In the wave function, the
bracketed numbers are the CI coefficients of a given electronic configuration.
Symbols $H$/$L$ denote HOMO/LUMO orbitals.}

\begin{tabular}{|c|c|c|c|}
\hline 
Peak	 & $E$ (eV) & $\mu^{2}$(a.u.) & Wave Function\tabularnewline
\hline
\hline 
IV & 4.1563 & 0.0919 & $\vert H\rightarrow L+3\rangle$(0.4902)\tabularnewline
\hline 
 &  &  & $\vert H\rightarrow L;H\rightarrow L+6\rangle$(0.4622)\tabularnewline
\hline 
 &  &  & $\vert H\rightarrow L;H\rightarrow L+5\rangle$(0.4070)\tabularnewline
\hline 
V & 5.0252 & 0.0171 & $\vert H-3\rightarrow L;H\rightarrow L+2\rangle$(0.3519)\tabularnewline
\hline 
 &  &  & $\vert H-2\rightarrow L;H\rightarrow L+2\rangle$(0.3322)\tabularnewline
\hline 
VII & 7.0917 & 0.3664 & $\vert H\rightarrow L+10\rangle$(0.5795)\tabularnewline
\hline 
 &  &  & $\vert H\rightarrow L;H\rightarrow L+9\rangle$(0.4872)\tabularnewline
\hline 
 &  &  & $\vert H\rightarrow L+1;H\rightarrow L+9\rangle$(0.3962)\tabularnewline
\hline 
IX & 8.2770 & 0.1232 & $\vert H-2\rightarrow L;H\rightarrow L+5\rangle$(0.3093)\tabularnewline
\hline 
 &  &  & $\vert H-2\rightarrow L;H\rightarrow L+3\rangle$(0.2999)\tabularnewline
\hline 
 & 8.3141 & 0.0905 & $\vert H\rightarrow L+1;H-1\rightarrow L+2\rangle$(0.3300)\tabularnewline
\hline 
 &  &  & $\vert H-1\rightarrow L+1;H\rightarrow L+4\rangle$(0.2080)\tabularnewline
\hline 
 &  &  & $\vert H-1\rightarrow L+3\rangle$(0.2043)\tabularnewline
\hline 
 & 8.3178 & 0.0974 & $\vert H-2\rightarrow L+1;H\rightarrow L+2\rangle$(0.2716)\tabularnewline
\hline 
 &  &  & $\vert H\rightarrow L+4;H-1\rightarrow L\rangle$(0.2525)\tabularnewline
\hline 
 &  &  & $\vert H-1\rightarrow L+1;H\rightarrow L+65\rangle$(0.2346)\tabularnewline
\hline 
 &  &  & $\vert H-2\rightarrow L;H\rightarrow L+4\rangle$(0.2121)\tabularnewline
\hline
\end{tabular}\label{Tab:cage-wavefunc-2}
\end{table}

\begin{table}
\caption{This table contains information pertinent to some of the peaks of
the INDO-MRSDCI optical absorption spectrum of the quasi-planar B$_{12}$,
as shown in Fig. \ref{fig:spectrum-indo-planar}. The symbols have
the same meaning as in the caption of table \ref{Tab:cage-wavefunc-2}. }

\begin{tabular}{|c|c|c|c|}
\hline 
Peak	 & E(eV) & $\mu^{2}$(a.u.) & Wave function\tabularnewline
\hline
\hline 
V & 10.7578 & 0.0279 & $\vert H\rightarrow L+8\rangle$(0.6230)\tabularnewline
\hline 
 &  &  & $\vert H-5\rightarrow L+2\rangle$(0.4956)\tabularnewline
\hline 
 &  &  & $\vert H\rightarrow L+1;H\rightarrow L+1\rangle$(0.2400)\tabularnewline
\hline 
 &  &  & $\vert H-5\rightarrow L\rangle$(0.1807)\tabularnewline
\hline 
VI & 11.2054 & 0.2187 & $\vert H-1\rightarrow L+9\rangle$(0.4144)\tabularnewline
\hline 
 &  &  & $\vert H-5\rightarrow L\rangle$(0.3539)\tabularnewline
\hline 
 &  &  & $\vert H\rightarrow L+10\rangle$(0.3255)\tabularnewline
\hline 
 &  &  & $\vert H-5\rightarrow L+2\rangle$(0.3049)\tabularnewline
\hline 
 &  &  & $\vert H\rightarrow L+8\rangle$(0.2973)\tabularnewline
\hline 
 &  &  & $\vert H\rightarrow L;H\rightarrow L\rangle$(0.1953)\tabularnewline
\hline 
VII & 12.3562 & 0.1598 & $\vert H\rightarrow L+10\rangle$(0.6469)\tabularnewline
\hline 
 &  &  & $\vert H-3\rightarrow L+3\rangle$(0.3408)\tabularnewline
\hline 
 &  &  & $\vert H\rightarrow L+12\rangle$(0.2985)\tabularnewline
\hline 
 &  &  & $\vert H\rightarrow L+13\rangle$(0.2299)\tabularnewline
\hline 
 &  &  & $\vert H\rightarrow L+8\rangle$(0.2095)\tabularnewline
\hline 
 &  &  & $\vert H-1\rightarrow L+11\rangle$(0.2063)\tabularnewline
\hline 
 &  &  & $\vert H-4\rightarrow L+5\rangle$(0.1792)\tabularnewline
\hline 
VIII & 13.1344 & 0.0300 & $\vert H-3\rightarrow L+3\rangle$(0.6665)\tabularnewline
\hline 
 &  &  & $\vert H-4\rightarrow L+5\rangle$(0.3184)\tabularnewline
\hline 
 &  &  & $\vert H\rightarrow L+13\rangle$(0.2582)\tabularnewline
\hline 
 &  &  & $\vert H-2\rightarrow L+2\rangle$(0.2586)\tabularnewline
\hline 
 &  &  & $\vert H-1\rightarrow L;H\rightarrow L+3\rangle$(0.2186)\tabularnewline
\hline 
 &  &  & $\vert H-4\rightarrow L+7\rangle$(0.1921)\tabularnewline
\hline 
 &  &  & $\vert H-4\rightarrow L;H\rightarrow L+2\rangle$(0.1574)\tabularnewline
\hline
\end{tabular}\label{Tab:plane-wave-func2}
\end{table}


\begin{thebibliography}{39}
\bibitem{ozturk}T. Ozturk, A. Demirbas, Energy Sources, Part A \textbf{29},
1415 (2007).

\bibitem{McKee}M. L. McKee, Z. Wang and P. v R. Schleyer, J. Am.
Chem. Soc. \textbf{122}, 4781 (2000).

\bibitem{Mutterties}E.L Mutterties, 'The Chemistry of Boron and its
Compounds', Wiely, New York, 1968.

\bibitem{lipscomb}W. L. Lipscomb, 'Boron Hydrides', W. A. Benjamin,
New York, 1963.

\bibitem{kbamba}G. Bambakidis and R. P. Wagner, J. Phys.Chem. Solids
\textbf{42},1023 (1981).

\bibitem{kawai}R. Kawai and J. H. Weare, J. Chem. Phys. \textbf{95},
1151 (1991).

\bibitem{kato}H. Kato and K. Yamashita, Chem. Phys. Lett. \textbf{190},
361(1992).

\bibitem{boustani-1}I. Boustani, Chem. Phys. Lett. \textbf{240},
135 (1995).

\bibitem{boustani-2}I. Boustani, Phys. Rev. B \textbf{55}, 16426
(1997).

\bibitem{fujimori}M. Fujimori and K. Kimura, J. Sol. St. Chem. \textbf{133},
310 (1997).

\bibitem{hayami}W. Hayami, Phys.Rev. \textbf{B} 60,1523 (1999).

\bibitem{zhai} H. J. Zhai, B. Kiran, J. LI, and L. S. Wang, Nature
Mat. \textbf{2}, 827 (2003).

\bibitem{b12-ionicity-prl}J. He, E. Wu, H. Wang, R. Liu, and Y. Tian,
Phys. Rev. Letts. \textbf{94}, 015504 (2005).

\bibitem{atis}M. Atis, C. Özdo\u{g}an, and Z. B. Güvenç, Int. J.
Quant. Chem. \textbf{107}, 729 (2007).

\bibitem{jemmis-prl}D. L. V. K. Prasad and E. D. Jemmis, Phys. Rev.
Lett. \textbf{100}, 165504 (2008).

\bibitem{perkin}C. L. Perkins, M. Trenary and T. Tanaka, Phys. Rev.
Lett. \textbf{77}, 4772 (1996).

\bibitem{hubert} H. Hubert, B. Devouard, L. A. J. Garvie, M. O'Keeffe,
P. R. Buseck, W. T. Petuskey and P. F. McMillan, Nature \textbf{391},
376 (1998).

\bibitem{reis-1}H. Reis and M. G. Papadopoulos, J. Comp. Chem. \textbf{20},
679 (1999).

\bibitem{reis-2}H. Reis, M. G. Papadopoulos, and I. Boustani, Int.
J. Quant. Chem. \textbf{78}, 131 (2000).

\bibitem{ayj}A. Abdurahman, A. Shukla, and G. Seifert, Phys. Rev.
B \textbf{66}, 155423 (2002).

\bibitem{xie}R. Xie, G. W. Bryant, J. Zhao, T. Kar and V. H. Smith
Jr., Phys. Rev. B \textbf{71}, 125422 (2005).

\bibitem{INDO}J. A. Pople, D. L. Beveridge and P. A. Dobosh, J. Chem.
Phys. \textbf{47}, 2026 (1967).

\bibitem{tddft-1}A. Zangwill and P. Soven, Phys. Rev. A \textbf{21},
1561 (1980).

\bibitem{tddft-2}E. Runge and E. Gross, Phys. Rev. Lett. \textbf{52},
997 (1984).

\bibitem{pople-book}For a review, see, \emph{e.g}., J. A. Pople and
D. L. Beveridge , Approximate Molecular Orbital Theory, McGraw-Hill
Publications, 1970.

\bibitem{CNDO}J.A. Pople, D.P. Santry, and G.A. Segal, J. Chem. Phys.
\textbf{43}, S 129 (1965); J.A. Pople and G.A. Segal, J. Chem. Phys.
\textbf{43},S136 (1965); J.A. Pople and G.A. Segal, J. Chem. Phys.
\textbf{44}, 3289 (1966).

\bibitem{CNDO-S}J. D. Bene and H. H. Jaffé, J. Chem. Phys. 48, 1807
(1968).

\bibitem{INDO-S1}J. E. Ridley and M. C. Zerner, Theor. Chim. Acta
\textbf{32}, 111 (1973).

\bibitem{INDO-S2}A. D. Bacon and M. C. Zerner, Theoret. Chim. Acta
\textbf{53,} 21 (1979).

\bibitem{INDO-S3}M. C. Zerner, G. H. Loew, R. F. Kirchner, and U.
T. Mueller-Westerhoff, J. Am. Chem. Soc. \textbf{102},589 (1980). 

\bibitem{INDOS-dye}M. Adachi and S. Nakamura, Dyes and Pigments \textbf{17},
287 (1991).

\bibitem{indo-paracetamol}A. S. El-Shahawy, S. M. Ahmed, and N. Kh.
Sayed, Spectrochimica Acta Part A \textbf{66}, 143 (2007).

\bibitem{shukla}S. Sahu and A. Shukla, Comp. Phys. Comm. \textbf{180},
724 (2009).

\bibitem{meld}We used modules sortin, cistar, rtsim and tmom of MELD,
a molecular electronic structure program from University of Indiana
with contributions from E. R. Davidson, L. McMurchie, S. Elbert, and
S. Langhoff.

\bibitem{mrsd-calc}See, \emph{e.g.}, P. Sony and A. Shukla, J. Chem.
Phys. \textbf{131}, 014302 (2009); P. Sony and A. Shukla, Phys. Rev.
B \textbf{75}, 155208 (2007); P. Sony and A. Shukla, Phys. Rev. B
\textbf{71}, 165204 (2005); A. Shukla, Phys. Rev. B \textbf{65}, 125204
(2002).

\bibitem{boron-band-gap}H. Werheit, A. Hausen, and H. Binnerbruck,
Phys. Stat. Solidi \textbf{42}, 733 (2006).

\bibitem{heer-rmp}W. A. de Heer, Rev. Mod. Phys. \textbf{65}, 611
(1993).

\bibitem{koutecky}J. Blanc, V. Bona\v{c}i\'{c}-Koutecký, M. Broyer,
J. Chevaleyre, Ph. Dugourd, J. Koutecký, C. Scheuch, J. P. Wolf, and
L. Wöste, J. Chem. Phys. \textbf{96}, 1793 (1992).

\bibitem{gaussian}Gaussian 03, Revision C.02, M. J. Frisch, G. W.
Trucks, H. B. Schlegel, G. E. Scuseria, M. A. Robb, J. R. Cheeseman,
J. A. Montgomery, Jr., T. Vreven, K. N. Kudin, J. C. Burant, J. M.
Millam, S. S. Iyengar, J. Tomasi, V. Barone, B. Mennucci, M. Cossi,
G. Scalmani, N. Rega, G. A. Petersson, H. Nakatsuji, M. Hada, M. Ehara,
K. Toyota, R. Fukuda, J. Hasegawa, M. Ishida, T. Nakajima, Y. Honda,
O. Kitao, H. Nakai, M. Klene, X. Li, J. E. Knox, H. P. Hratchian,
J. B. Cross, V. Bakken, C. Adamo, J. Jaramillo, R. Gomperts, R. E.
Stratmann, O. Yazyev, A. J. Austin, R. Cammi, C. Pomelli, J. W. Ochterski,
P. Y. Ayala, K. Morokuma, G. A. Voth, P. Salvador, J. J. Dannenberg,
V. G. Zakrzewski, S. Dapprich, A. D. Daniels, M. C. Strain, O. Farkas,
D. K. Malick, A. D. Rabuck, K. Raghavachari, J. B. Foresman, J. V.
Ortiz, Q. Cui, A. G. Baboul, S. Clifford, J. Cioslowski, B. B. Stefanov,
G. Liu, A. Liashenko, P. Piskorz, I. Komaromi, R. L. Martin, D. J.
Fox, T. Keith, M. A. Al-Laham, C. Y. Peng, A. Nanayakkara, M. Challacombe,
P. M. W. Gill, B. Johnson, W. Chen, M. W. Wong, C. Gonzalez, and J.
A. Pople, Gaussian, Inc., Wallingford CT, 2004.

\end{thebibliography}
\end{document}